\documentclass[lettersize,journal]{IEEEtran}
\usepackage{amsmath,amsfonts,amsthm,amssymb,amsbsy}
\usepackage{algorithmic}
\usepackage{algorithm}
\usepackage{array}
\usepackage[caption=false,font=normalsize]{subfig}
\usepackage{textcomp}
\usepackage{stfloats}
\usepackage{url}
\usepackage{verbatim}
\usepackage{graphicx}
\usepackage{cite}
\newtheorem{assumption}{Assumption}

\newtheorem{theorem}{Theorem}
\usepackage{xcolor}

\begin{document}

\title{Semi-Blind Channel Estimation for Dynamic NTN Systems via Spiked Random Matrix Theory
}
\author{
Xue Zhang, {\em Graduate Student Member, IEEE}, Abla Kammoun, {\em Member, IEEE}, \\ and Mohamed-Slim Alouini, {\em Fellow, IEEE}
\thanks{The authors are with Computer, Electrical and Mathematical Sciences
and Engineering (CEMSE) Division, Department of Electrical and Computer
Engineering, King Abdullah University of Science and Technology (KAUST), Thuwal 23955-6900, Saudi Arabia. (e-mail: xue.zhang@kaust.edu.sa; abla.kammoun@kaust.edu.sa; slim.alouini@kaust.edu.sa). (Corresponding author: Xue Zhang). 
}
\vspace{-8mm}
}

\maketitle

\vspace{-0.8cm}

\begin{abstract}
Semi-blind channel estimation offers an attractive tradeoff between pilot overhead and estimation accuracy in large-scale wireless systems. However, reliable channel acquisition becomes particularly challenging in highly dynamic environments such as non-terrestrial networks (NTNs), where rapidly varying channels and high system dimensionality significantly degrade the performance of conventional covariance-based estimators due to sampling noise. In this paper, we propose a robust semi-blind channel estimation framework for multi-user uplink systems operating in NTN systems. The proposed approach introduces an optimally regularized least-squares formulation that balances training-based information and blind subspace structure. By exploiting the spiked covariance model within a random matrix theory (RMT) framework, we derive a closed-form characterization of the resulting channel mean-squared error and obtain an analytically tractable design of the optimal regularization parameter. The resulting estimator is computationally efficient and particularly well suited to high-dimensional regimes. Simulation results under realistic Third Generation Partnership Project (3GPP) NTN channel models demonstrate substantial performance improvements over conventional semi-blind and training-based estimators.
\end{abstract}


\section{Introduction}
\label{sec_introdution}
Future wireless systems, including sixth-generation (6G) networks and non-terrestrial networks (NTNs), are expected to operate under highly dynamic propagation environments, where platform mobility and propagation characteristics can considerably shorten the effective channel coherence interval, making reliable channel acquisition particularly challenging.~\cite{zhang2024joint,cao2025survey,zhang2026design}. To address these challenges, significant research efforts have been devoted to the design of advanced waveforms capable of adapting to dynamic wireless channels. Representative examples include delay–Doppler domain waveforms such as orthogonal time frequency space (OTFS) and orthogonal delay–Doppler division multiplexing (ODDM), chirp-based waveforms such as orthogonal chirp division multiplexing (OCDM) and affine frequency division multiplexing (AFDM), as well as enhanced orthogonal frequency division multiplexing (OFDM) variants~\cite{deng2025unifying}. These waveform designs aim to exploit the inherent structure of dynamic channels and improve communication robustness in rapidly varying environments.

\par 

The performance of such advanced waveform systems critically depends on the availability of accurate channel state information (CSI) at the receiver. In practice, CSI is obtained through channel estimation, which plays a fundamental role in enabling reliable detection and interference mitigation in multiple-input multiple-output (MIMO) systems~\cite{wang2007performance,zhou2022channel}. However, reliable channel acquisition becomes particularly challenging in dynamic wireless environments and large-scale MIMO systems due to rapidly varying channels and the high dimensionality of the signal space. Channel estimation techniques can generally be categorized into blind and pilot-aided approaches. Blind channel estimation relies solely on received data signals and exploits the statistical properties of the transmitted symbols to recover the channel~\cite{noh2014new}. Although blind methods can improve spectral efficiency by avoiding pilot overhead, they typically suffer from intrinsic ambiguities and high computational complexity~\cite{tong1994blind}. In contrast, pilot-aided channel estimation inserts predefined training sequences, commonly referred to as pilot symbols, into the transmitted frame to facilitate channel acquisition~\cite{hassibi2003much}. While this approach is widely adopted due to its robustness and relatively low complexity, the use of pilot symbols inevitably consumes valuable time-frequency resources and reduces spectral efficiency.

\par

To balance the advantages and limitations of blind and pilot-aided methods, semi-blind channel estimation has been proposed, which jointly exploits pilot symbols and unknown data symbols to improve estimation accuracy while reducing pilot overhead~\cite{srinivas2019iterative}. Compared with purely blind approaches, semi-blind methods leverage a limited number of pilot symbols to resolve channel ambiguities and enhance estimation reliability~\cite{aldana2003channel}. This property is particularly attractive in large-scale MIMO systems, where channel estimation becomes increasingly challenging due to the large number of parameters involved. Consequently, semi-blind channel estimation emerges as a promising technique for improving channel acquisition performance without increasing pilot overhead.


\par

A substantial body of literature has explored semi-blind channel estimation schemes in multi-user MIMO systems. Among these approaches, maximum-likelihood (ML) estimation techniques, particularly those based on expectation-maximization (EM) algorithms, have attracted significant attention. In~\cite{aldana2003channel}, a frequency-domain EM algorithm was proposed for estimating both blind and semi-blind channels for each user in an underdetermined multiple-input single-output (MISO) system. Similarly, the work in \cite{nayebi2017semi} investigated two EM-based schemes for semi-blind channel estimation, assuming Gaussian signaling for the unknown data symbols. To further enhance computational efficiency,~\cite{al2021semi} employed eigenvalue decomposition to accelerate the EM algorithm with a Gaussian prior and leveraged the discrete prior of data symbols to derive a tractable version of the EM procedure. In addition to EM-based methods, iterative techniques have also been investigated to compute ML channel estimators. In this context,~\cite{abuthinien2008semi} proposed an iterative two-level optimization loop for jointly estimating channel coefficients and training sequences in MIMO systems. Beyond ML estimators, another class of methods extends blind estimation techniques by incorporating training sequence information to resolve the inherent matrix ambiguities of blind approaches. For instance,~\cite{lawal2023semi} introduced a semi-blind structured signal subspace algorithm to simultaneously reduce computational complexity and minimize the required training sequence size. Similarly, \cite{rekik2024fast} developed a blind and semi-blind subspace-based algorithm designed to reduce the computational cost, accelerate convergence, and achieve accurate channel estimates with a smaller sample size. 

\par

In parallel with these developments, the authors in~\cite{buchoux2000performance} studied a simpler semi-blind channel estimation approach initially suggested in~\cite{gorokhov1997semi}, which linearly combines the least-squares (LS) fit on the training sequence with the blind subspace criterion introduced in~\cite{moulines2002subspace}. The resulting estimation criterion in~\cite{gorokhov1997semi} is quadratic, eliminating the need for iterative optimization, and can be interpreted as a regularized least-squares formulation where the statistical structure of the observations acts as a regularization constraint. Motivated by these advantages, the authors in~\cite{buchoux2000performance} applied the approach to a single-input multiple-output (SIMO) system, deriving a semi-blind channel estimator where the blind subspace criterion serves as the regularization term. Furthermore, an asymptotic analysis was conducted to optimally tune the balance between the blind criterion and the LS fit based on the system design parameters. However, while effective for SIMO systems, this approach does not directly extend to massive MIMO systems under high-dimensional settings, where the large number of antennas, and the associated dimensionality challenges necessitate the development of new estimation strategies.

\par

In this paper, we investigate the problem of semi-blind channel estimation for uplink multi-user MIMO systems operating in dynamic wireless environments, with particular emphasis on NTN scenarios. Specifically, we consider a high-dimensional regime in which the number of antennas, the number of pilot symbols, and the total length of the data transmission block grow large with comparable orders. This regime is practically relevant, as many real-world massive MIMO systems operate under long coherence intervals where the channel remains approximately constant within one coherence block~\cite{bjornson2017massive}. However, in such high-dimensional settings, conventional covariance-based channel estimation techniques often suffer from significant performance degradation due to sampling noise and scalability limitations. To address this, we leverage the statistical structure of the received data covariance matrix, which aligns with the well-known "spiked model" from random matrix theory (RMT). The spiked model assumes that the population covariance matrix is a low-rank perturbation of a scaled identity matrix, with the scaling factor corresponding to the noise variance, denoted by $\sigma_v^2$. In this model, the dominant eigenmodes capture the signal subspace, while the remaining eigenmodes, which have eigenvalues equal to $\sigma_v^2$, correspond to the noise subspace. Such structure enables us to develop a robust semi-blind channel estimation framework. In particular, under the RMT asymptotic regime, we derive an asymptotic expression for the MSE of the semi-blind estimator. By analyzing this expression, we identify the optimal value of the regularization parameter and construct a consistent estimator for it. This, in turn, enables the design of an efficient semi-blind channel estimator whose performance is asymptotically optimal among all choices of the regularization parameter.

\par

To the best of our knowledge, the use of spiked covariance models from RMT to design semi-blind channel estimation algorithms for high-dimensional dynamic wireless systems has not been previously investigated. While related works exist, such as~\cite{johnstone2009consistency}, which studied optimized spiked-model-based covariance estimators under rotationally invariant loss functions, and~\cite{yang2018high}, which applied similar concepts to the minimum variance distortionless response (MVDR) beamforming problem, these approaches are not directly applicable to the semi-blind channel estimation problem considered in this work, particularly in large-scale MIMO systems operating in dynamic wireless environments. Motivated by these observations, we develop a new and easily implementable semi-blind channel estimation framework tailored for high-dimensional multi-user MIMO systems, with particular relevance to emerging NTN scenarios and advanced waveform-based communication systems. Our main contributions are summarized as follows:
\begin{itemize}
    \item We propose a semi-blind channel estimator by minimizing a regularized least-squares cost function that combines training sequence-based information with a blind subspace criterion, and derive a closed-form expression for the resulting channel mean squared error (MSE) accordingly.   
    \item We leverage spiked models from random matrix theory to analytically optimize the regularization parameter that balances the contributions of the training-based and blind components, thereby minimizing the MSE and yielding improved channel coefficient estimates.  
    \item We perform numerical simulations in realistic Third Generation Partnership Project (3GPP) NTN environments to validate the accuracy of the proposed asymptotic expressions and to demonstrate the advantages of the semi-blind channel estimator over classical methods in dynamic wireless environments.
\end{itemize}

The remainder of this paper is organized as follows. Section~\ref{sec_model} introduces the system model. Section~\ref{sec_criterion} presents the regularized semi-blind channel estimator and derives the corresponding MSE expression. In Section~\ref{sec_channel_design}, we employ the spiked model from random matrix theory to optimize the regularization parameter for MSE minimization and derive the corresponding optimal channel estimates. Section~\ref{sec_sim} provides simulation results under realistic NTN channel models to validate the proposed approach. Finally, Section~\ref{sec_con} concludes the paper.

\textit{Notations}: The vectors (matrices) are denoted by lower-case (upper-case) boldface characters. $\mathbf{I}_{M}$ denotes the $M\times M$ identity matrix, and $\mathbf{0}$ stands for a zero matrix with appropriate dimension. $\mathbb{E}\{\cdot\}$ represents the expectation operator. The trace of $\mathbf{A}$ are denoted by $\mathrm{tr}(\mathbf{A})$. Superscripts $H$ denote the conjugate transpose. $|\cdot|$ denotes the absolute value. $\|\cdot\|_2$ stand for the $l_2$ norm. $\mathbf{A}\succeq\mathbf{0}$ is equivalent to $\mathbf{A}$ being positive semidefinite. $\delta_x$ denotes the Dirac measure at point $x$. For a sequence of random variables $({X}_{n})_{n\in\mathbb{N}}$ and a random variable $\mathbf{X}$, we write 
\begin{align}
    \mathbf{X}_n\xrightarrow[N\rightarrow+\infty]{a.s.}\mathbf{X}, \,\,\text{and}\,\,\mathbf{X}_n\xrightarrow[N\rightarrow+\infty]{\mathcal{D}}\mathbf{X},
\end{align}
to indicate that $\mathbf{X}_n$ converges to $\mathbf{X}$ almost surely and in distribution, respectively. $\mathcal{O}(\cdot)$ represents the convolution Big-O notation.


\section{System Model}
\label{sec_model}

We consider a single-cell uplink multi-user MIMO system where a base station (BS) equipped with $M$ antennas communicates with $K$ single-antenna users, with $K<M$. Such a setting naturally arises in emerging wireless systems operating over highly dynamic propagation environments, including NTNs and high-mobility communication scenarios. The channel matrix between the BS and the users is denoted by $\mathbf{G}\in\mathbb{C}^{M\times K}$. We consider a multicarrier transmission framework, which encompasses a wide class of advanced waveform designs proposed for dynamic wireless channels. In particular, OFDM can be viewed as a representative case, and we adopt it here for analytical tractability. For each subcarrier, we use a narrowband (flat) fading approximation so that the channel on that subcarrier is fully characterized by $\mathbf{G}$. We adopt a block-fading model in which $\mathbf{G}$ remains constant over a coherence block of $N$ symbols and changes independently from block to block. The uplink transmission within each coherence block consists of $L$ pilot symbols followed by $(N-L)$ data symbols. The received signal at the BS on a given subcarrier at time $n$ is modeled as~\cite{nayebi2017semi}
\begin{equation}
    \mathbf{y}[n] = \mathbf{G}\mathbf{s}[n] + \mathbf{v}[n],
\end{equation}
where for $n=0,\ldots,L-1$, the transmitted vectors $\mathbf{s}(n)=[s_1(n),\ldots,s_K(n)]^T$ correspond to known pilot symbols, while for $n=L,\ldots,N-1$ they represent unknown data symbols with unit average power, i.e., $\mathbb{E}\{\mathbf{s}(n)\mathbf{s}(n)^H\}=P_s\mathbf{I}_K$. The noise vectors are modeled as $\mathbf{v}(n)\sim\mathcal{CN}(\mathbf{0},\sigma_v^2\mathbf{I}_M)$. Let $\mathbf{S}_p=[\mathbf{s}(0),\ldots,\mathbf{s}(L-1)]$ and $\mathbf{S}_d=[\mathbf{s}(L),\ldots,\mathbf{s}(N-1)]$ denote the pilot and data symbol matrices within the coherence block, respectively. The complete transmitted symbol matrix is therefore given by $\mathbf{S}=[\mathbf{S}_p,\mathbf{S}_d]$, and the received signal and noise matrices decompose as $\mathbf{Y}=[\mathbf{Y}_p,\mathbf{Y}_d]$ and $\mathbf{V}=[\mathbf{V}_p,\mathbf{V}_d]$, respectively.

The following assumption formalizes the statistical properties of the pilot and data symbols used in our model.
\begin{assumption}\label{ass1_symbol}
We assume that the pilot sequences are orthogonal such that $\mathbf{S}_p\mathbf{S}_p^H=aL\mathbf{I}_K$, where $a$ denotes the average pilot transmit power per user. The unknown symbol vectors $\mathbf{s}_n$, $n=L,\cdots,N-1$, are independent with independent entries and covariance $P_s\mathbf{I}_K$.
\end{assumption}

\par

\section{Semi-blind Criterion}
\label{sec_criterion}
This section develops the semi-blind channel estimation criterion used for receiver design in dynamic multi-user MIMO systems. In the case of unknown data symbols, the covariance matrix of the received signal $\mathbf{y}(n)$ for $n=L, \ldots, N-1$ is defined as
\begin{align}
    \mathbf{R} =& \mathbb{E}\{\mathbf{y}(n)\mathbf{y}(n)^H\} = P_s\mathbf{G}\mathbf{G}^H + \sigma_v^2\mathbf{I}_M. 
\end{align}
Applying eigen-decomposition to $\mathbf{R}$ yields
\begin{align}
    \mathbf{R}=\mathbf{U}_s\pmb{\Lambda}_s\mathbf{U}_s^H + \sigma_v^2\mathbf{U}_n\mathbf{U}_n^H,
\end{align}
where $\pmb{\Lambda}_s$ is a $K\times K$ diagonal matrix containing the largest eigenvalues of $\mathbf{R}$, $\mathbf{U}_s$ is a $M\times K$ matrix that contains the eigenvectors corresponding to the $K$ largest eigenvalues, and $\mathbf{U}_n$ is a $M\times (M-K)$ matrix that contains the eigenvectors associated with the smallest eigenvalue $\sigma_v^2$. Thus the eigenvalues and eigenvectors of $\mathbf{R}$ naturally split into two orthogonal subspaces: the signal subspace spanned by $\mathbf{U}_s$, and the noise subspace spanned by $\mathbf{U}_n$. A key property exploited by subspace-based identification algorithms~\cite{johnson2008music,mestre2008modified,vallet2015performance,zhang2022doa} is that any vector belonging to the signal subspace is orthogonal to the noise subspace. In our case, we have 
\begin{align}\label{orth_U}
    \mathbf{g}_k^H\mathbf{U}_n = 0,
\end{align}
for any $k=1,\ldots,K$. Consequently, the channel vectors ${\mathbf{g}_k}$, $k=1,\ldots,K$, lie among the directions that minimize the pseudo-spectrum 
$P(\mathbf{g},\mathbf{U}_n)=\mathbf{g}^H\mathbf{U}_n\mathbf{U}_n^H\mathbf{g}$.

Since $\mathbf{U}_n$ is not known in practice, it must be estimated from the data matrix $\mathbf{Y}_d$ using its sample covariance matrix:
\begin{align}
    \hat{\mathbf{R}} = \frac{1}{N-L}\mathbf{Y}_d\mathbf{Y}_d^H. 
\end{align}
Performing eigen-decomposition of $\hat{\mathbf{R}}$, we can obtain the estimated $\hat{\mathbf{U}}_s$ and $\hat{\mathbf{U}}_n$ by selecting the eigenvectors corresponding to the $K$ largest and $(M-K)$ smallest eigenvalues, respectively. Note that the impact of estimation errors can be severe, particularly when the number of unknown sample available for estimation $(N-L)$ is not substantially larger than the observation dimensionality $M$. To estimate the channel vectors, we search for the most significant minima of the estimated pseudo-spectrum: $P(\mathbf{g}_k,\hat{\mathbf{U}}_n)=\mathbf{g}_k^H\hat{\mathbf{U}}_n\hat{\mathbf{U}}_n^H\mathbf{g}_k$~\cite{vallet2015performance}, i.e., 
\begin{align}
    \hat{\mathbf{g}}_k = \arg \min_{\|\mathbf{g}_k\|=1} P(\mathbf{g}_k,\hat{\mathbf{U}}_n).
\end{align}
Combining all $K$ channel vectors, the blind subspace estimator is given by 
\begin{align}\label{G_blind}
    \hat{\mathbf{G}}_{\mathrm{blind}} =& \arg\min_{\|\mathbf{g}_k\|=1} \sum_{k=1}^KP(\mathbf{g}_k,\hat{\mathbf{U}}_n) \notag\\
    =& \arg\min\mathrm{tr}(\mathbf{G}^H\hat{\mathbf{U}}_n\hat{\mathbf{U}}_n^H\mathbf{G}). 
\end{align}

On the other hand, when the known pilot symbols are assumed to satisfy $\mathbf{S}_p\mathbf{S}_p^H=aL\mathbf{I}_K$, we have
\begin{align}
    \mathbf{Y}_p\mathbf{S}_p^H = \mathbf{G}\mathbf{S}_p\mathbf{S}_p^H + \mathbf{V}_p\mathbf{S}_p^H,
\end{align}
leading to the exact channel expression: 
\begin{align}\label{channel vector}
    \mathbf{G} = \frac{1}{aL}\mathbf{Y}_p\mathbf{S}_p^H - \frac{1}{a L}\mathbf{V}_p\mathbf{S}_p^H, 
\end{align}
thus the training-based estimator is given by 
\begin{align}\label{G_training}
    \hat{\mathbf{G}}_{\mathrm{training}} =& \arg \min \left\|\mathbf{G} - \frac{1}{a L}\mathbf{Y}_p\mathbf{S}_p^H\right\|_F^2 \notag\\
    =& \frac{1}{a L}\mathbf{Y}_p\mathbf{S}_p^H. 
\end{align}

A natural way to construct a semi-blind estimation criterion is to linearly combine the training-based criterion~(\ref{G_training}) with the blind subspace criterion  to obtain~\cite{buchoux2000performance} 
\begin{align}\label{C_SB}
    C_{SB}(\mathbf{G}) = \lambda\left\|\mathbf{G} - \frac{1}{a L}\mathbf{Y}_p\mathbf{S}_p^H\right\|_F^2 + (1-\lambda)\mathrm{tr}(\mathbf{G}^H\hat{\mathbf{U}}_n\hat{\mathbf{U}}_n^H\mathbf{G}),
\end{align}
where $\lambda\in[0,1]$ is a regularization parameter that balances the influence of the training data and the blind information.  To minimize (\ref{C_SB}), let 
\begin{align}
    \frac{\partial C_{SB}(\mathbf{G})}{\partial\mathbf{G}} =& 2\lambda\left(\mathbf{G}-\frac{1}{aL}\mathbf{Y}_p\mathbf{S}_p^H\right) \notag\\
    &+ 2(1-\lambda)\hat{\mathbf{U}}_n\hat{\mathbf{U}}_n^H\mathbf{G} \triangleq \pmb{0},
\end{align}
then we can obtain channel coefficients estimates as   
\begin{align}\label{estimated channel vector}
    \hat{\mathbf{G}} = \left(\frac{1-\lambda}{\lambda}\hat{\mathbf{U}}_n\hat{\mathbf{U}}_n^H+\mathbf{I}_M\right)^{-1}\frac{1}{aL}\mathbf{Y}_p\mathbf{S}_p^H. 
\end{align}
From (\ref{channel vector}) and (\ref{estimated channel vector}), the estimation error is given by 
\begin{align}
    \hat{\mathbf{G}} - \mathbf{G} =& \left(\frac{1-\lambda}{\lambda}\hat{\mathbf{U}}_n\hat{\mathbf{U}}_n^H+\mathbf{I}_M\right)^{-1}\left(\mathbf{G}+\frac{1}{a L}\mathbf{V}_p\mathbf{S}_p^H\right) - \mathbf{G} \notag \\
    =& -\frac{1-\lambda}{\lambda}\hat{\mathbf{U}}_n\hat{\mathbf{U}}_n^H\left(\frac{1-\lambda}{\lambda}\hat{\mathbf{U}}_n\hat{\mathbf{U}}_n^H+\mathbf{I}_M\right)^{-1}\mathbf{G} \notag \\
    &+ \left(\frac{1-\lambda}{\lambda}\hat{\mathbf{U}}_n\hat{\mathbf{U}}_n^H+\mathbf{I}_M\right)^{-1}\frac{1}{a L}\mathbf{V}_p\mathbf{S}_p^H,
\end{align}
then the corresponding MSE is defined as 
\begin{align}\label{MSE_original}
    \mathrm{MSE}(\lambda) =& \frac{1}{K}\mathbb{E}\{\|\hat{\mathbf{G}} - \mathbf{G}\|_F^2\} \notag\\
    =& \frac{(1-\lambda)^2}{\lambda^2K}\mathrm{tr}\left(\mathbf{G}^H\left(\frac{1-\lambda}{\lambda}\hat{\mathbf{U}}_n\hat{\mathbf{U}}_n^H+\mathbf{I}_M\right)^{-1}\hat{\mathbf{U}}_n\right. \notag\\
    &\left.\cdot\hat{\mathbf{U}}_n^H\left(\frac{1-\lambda}{\lambda}\hat{\mathbf{U}}_n\hat{\mathbf{U}}_n^H+\mathbf{I}_M\right)^{-1}\mathbf{G}\right) \notag\\
    &+ \frac{\sigma_v^2}{aL}\mathrm{tr}\left(\left(\frac{1-\lambda}{\lambda}\hat{\mathbf{U}}_n\hat{\mathbf{U}}_n^H+\mathbf{I}_M\right)^{-2}\right). 
\end{align}
Note that $\hat{\mathbf{U}}_n$ and $\hat{\mathbf{U}}_s$ are orthogonal projection matrices and satisfy the identity $\hat{\mathbf{U}}_s\hat{\mathbf{U}}_s^H+\hat{\mathbf{U}}_n\hat{\mathbf{U}}_n^H=\mathbf{I}_M$, (\ref{MSE_original}) can be further rewritten as (\ref{MSE}). 

\begin{figure*}[t]
\begin{align}\label{MSE}
    \mathrm{MSE}(\lambda) =& \frac{(1-\lambda)^2}{\lambda^2K}\mathrm{tr}\left(\mathbf{G}^H\left(\hat{\mathbf{U}}_s\hat{\mathbf{U}}_s^H+\lambda\hat{\mathbf{U}}_n\hat{\mathbf{U}}_n^H\right)(\mathbf{I}_M-\hat{\mathbf{U}}_s\hat{\mathbf{U}}_s^H)\left(\hat{\mathbf{U}}_s\hat{\mathbf{U}}_s^H+\lambda\hat{\mathbf{U}}_n\hat{\mathbf{U}}_n^H\right)\mathbf{G}\right) + \frac{\sigma_v^2}{aL}\mathrm{tr}\Big(\Big(\hat{\mathbf{U}}_s\hat{\mathbf{U}}_s^H+\lambda\hat{\mathbf{U}}_n\hat{\mathbf{U}}_n^H\Big)^{2}\Big) \notag\\
    =& \frac{(1-\lambda)^2}{K}\mathrm{tr}\left(\mathbf{G}^H\left(\mathbf{I}_M+\frac{1-\lambda}{\lambda}\hat{\mathbf{U}}_s\hat{\mathbf{U}}_s^H\right)^2\mathbf{G}\right) + \frac{\sigma_v^2\lambda^2}{aL}\mathrm{tr}\left(\left(\mathbf{I}_M+\frac{1-\lambda}{\lambda}\hat{\mathbf{U}}_s\hat{\mathbf{U}}_s^H\right)^{2}\right) \notag\\ 
    &- \frac{(1-\lambda)^2}{K}\mathrm{tr}\left(\mathbf{G}^H\left(\mathbf{I}_M+\frac{1-\lambda}{\lambda}\hat{\mathbf{U}}_s\hat{\mathbf{U}}_s^H\right)\hat{\mathbf{U}}_s\hat{\mathbf{U}}_s^H\left(\mathbf{I}_M+\frac{1-\lambda}{\lambda}\hat{\mathbf{U}}_s\hat{\mathbf{U}}_s^H\right)\mathbf{G}\right) \notag\\
    =& \frac{(1-\lambda)^2}{K}\mathrm{tr}\left(\mathbf{G}^H(\mathbf{I}_M-\hat{\mathbf{U}}_s\hat{\mathbf{U}}_s^H)\mathbf{G}\right) + \frac{\sigma_v^2(\lambda^2M+(1-\lambda^2)K)}{aL} \notag\\
    =& \frac{(1-\lambda)^2}{K}\sum_{k=1}^K\mathbf{g}_k^H\mathbf{g}_k - \frac{(1-\lambda)^2}{K}\sum_{k=1}^K\sum_{i=1}^K\mathbf{g}_k^H\hat{\mathbf{u}}_{si}\hat{\mathbf{u}}_{si}^H\mathbf{g}_k + \frac{\sigma_v^2(\lambda^2M+(1-\lambda^2)K)}{aL}.
\end{align}
\hrulefill
\vspace*{10pt}
\end{figure*}

It is well known that the estimated channel matrix $\hat{\mathbf{G}}$ can exhibit substantial errors compared to the true channel matrix $\mathbf{G}$. A primary source of this error lies in the estimation of the noise subspace $\hat{\mathbf{U}}_n$, which becomes particularly unreliable when the number of antennas $M$ and the number of unknown data samples $(N - L)$ are of the same order of magnitude. This situation frequently arises in large-scale wireless systems operating in dynamic propagation environments, where reliable covariance estimation becomes challenging. In the considered massive MIMO system, both $M$ and $(N - L)$ are assumed to be reasonably large, while the number of users $K$ is relatively small and fixed. Under these conditions, the covariance matrix $\mathbf{R}$ can be characterized as a low-rank (rank-$K$) perturbation of the identity matrix. From the perspective of RMT, this configuration corresponds to the so-called spiked covariance model. This model has been the subject of extensive recent research~\cite{baik2005phase,baik2006eigenvalues,paul2007asymptotics,el2007tracy,nadler2008finite}, which has provided valuable insights into the statistical behavior of sample covariance matrices with low-rank perturbations. Building on these insights, we propose to incorporate the theoretical properties of spiked covariance models into the design of a new semi-blind channel estimation strategy. To the best of our knowledge, this approach has not been previously explored in this specific context, representing a key contribution of our work. Further details are provided in Section~\ref{sec_channel_design}.

\section{Optimized Channel Matrix Design}
\label{sec_channel_design}
In high-mobility and non-stationary propagation environments envisioned for future 6G/NTN systems and advanced waveform receivers, reliable channel acquisition is limited by the accuracy of sample covariance estimation in high dimensions. In this section, we leverage the known spiked covariance structure $\mathbf{R}$ to propose an optimized regularization parameter $\hat{\lambda}^\ast$, based on which an optimized channel matrix $\hat{\mathbf{G}}^\ast$ is designed to minimize the objective function in (\ref{MSE_original}).

\subsection{Spiked Covariance Models}
This subsection briefly reviews the main RMT results associated with spiked covariance models that are relevant to the subsequent analysis and will be repeatedly used throughout the remainder of the paper.

The covariance matrix $\mathbf{R}$ using eigen-decomposition can be rewritten as
\begin{align}
    \mathbf{R} = \sigma_v^2\left(\sum_{k=1}^Kt_k\mathbf{u}_{sk}\mathbf{u}_{sk}^H + \mathbf{I}_M\right),
\end{align}
which has eigenvalues $\sigma_v^2(t_1+1),\ldots,\sigma_v^2(t_K+1),\underbrace{\sigma_v^2,\ldots,\sigma_v^2}_{M-K\ \text{times}}$. Here, $t_k>0$ for $k=1,\ldots, K$, $t_1\sigma_v^2+\sigma_v^2,\ldots,t_K\sigma_v^2+\sigma_v^2$ are referred to as "spiked" eigenvalues, and the associated eigenvectors $\mathbf{u}_{s1},\ldots,\mathbf{u}_{sK}$ span the signal subspace $\mathbf{U}_s$. 

Our proposed estimator is designed to exploit this spectral structure by manipulating the eigenvalue distribution of the sample covariance matrix. Specifically, $\hat{\mathbf{R}}$ via eigen-decomposition can be given by 
\begin{align}\label{hat_R_evd}
    \hat{\mathbf{R}} = \sum_{i=1}^{K}\eta_i\hat{\mathbf{u}}_{si}\hat{\mathbf{u}}_{si}^H + \sum_{i=K+1}^{M}\eta_i\hat{\mathbf{u}}_{ni}\hat{\mathbf{u}}_{ni}^H,
\end{align}
where $\eta_1\geq \cdots\geq \eta_K$ denotes the $K$ largest eigenvalue, while $\eta_{K+1}\geq \cdots\geq \eta_M$ denote the last $M-K$ largest eigenvalue. Here, the vectors $\hat{\mathbf{u}}_{si}$ for $i=1,\ldots,K$ and $\hat{\mathbf{u}}_{ni}$ for $i=K+1,\ldots,M$ represent the columns of $\hat{\mathbf{U}}_s$ and $\hat{\mathbf{U}}_n$, respectively.

Assuming that $K$ remains fixed while all other dimensions $M$, $N$, and $L$ grow large at the same rate, 
the sample covariance matrix $\hat{\mathbf{R}}$ falls into the framework of a spiked random covariance model. 
In this regime, the empirical eigenvalue distribution consists of a {bulk} formed by the smallest $M-K$ eigenvalues, 
together with at most $K$ outlier eigenvalues (spikes) that separate from the bulk. 
The eigenvalue bulk follows the classical Marchenko-Pastur law, whose support and shape are determined by the noise variance 
and the dimensional ratio. 
The remaining eigenvalues correspond to the informative signal components and may escape from the bulk. 
Their behavior is governed by a well-known phase transition phenomenon: 
only spikes whose population strength exceeds a critical threshold emerge from the bulk, 
while weaker spikes remain embedded within it ~\cite{couillet2011random,baik2006eigenvalues,vallet2015performance,yang2018high}. To make these ideas precise, we introduce the following assumption, which will underlie our study.

\begin{assumption}\label{ass1}
    We assume that the number $K$ of users is fixed, while the number $M$ of antennas, the number $L$ of pilot symbols, and the length $N$ of the whole transmission block go to infinity at the same rate, that $M, N, L, N-L\rightarrow+\infty$ with $M/N\rightarrow \alpha$, $L/N\rightarrow \beta$, $M/(N-L)=c_M\rightarrow c$ for certain $\alpha,\beta,c>0$. The number of spikes $K$ is fixed, independently of $M$ and $N-L$, while $t_1>\ldots>t_K$ with $t_K>\sqrt{c}$ for all large $M$. 
\end{assumption}

Throughout the paper, the notation $\xrightarrow[N\to\infty]{}$ denotes convergence under this regime, with $N \to \infty$ standing for the asymptotic setting introduced above.

Let $\hat{\mu}_{\mathbf{R}}(\lambda)$ be the empirical spectral distribution (ESD) of eigenvalues of the sample covariance matrix $\hat{\mathbf{R}}$, defined as 
\begin{align}
    \hat{\mu}_{\mathbf{R}}(\lambda) = \frac{1}{M}\sum_{i=1}^M\delta_{\eta_i},
\end{align}
which can be alternatively characterized through its Stieltjes transform defined as~\cite{vallet2015performance} 
\begin{align}
    \hat{m}_{\mathbf{R}}(z) =& \int_{\mathbb{R}}\frac{1}{\lambda-z}d\mu_{\mathbf{R}}(\lambda) = \frac{1}{M}\sum_{i=1}^M\frac{1}{\eta_i-z} \notag \\
        =& \frac{1}{M}\mathrm{tr}\left(\hat{\mathbf{R}}-z\mathbf{I}_M\right)^{-1}.
\end{align}

\begin{figure}[t]
\centering
\includegraphics[width=1.0\linewidth]{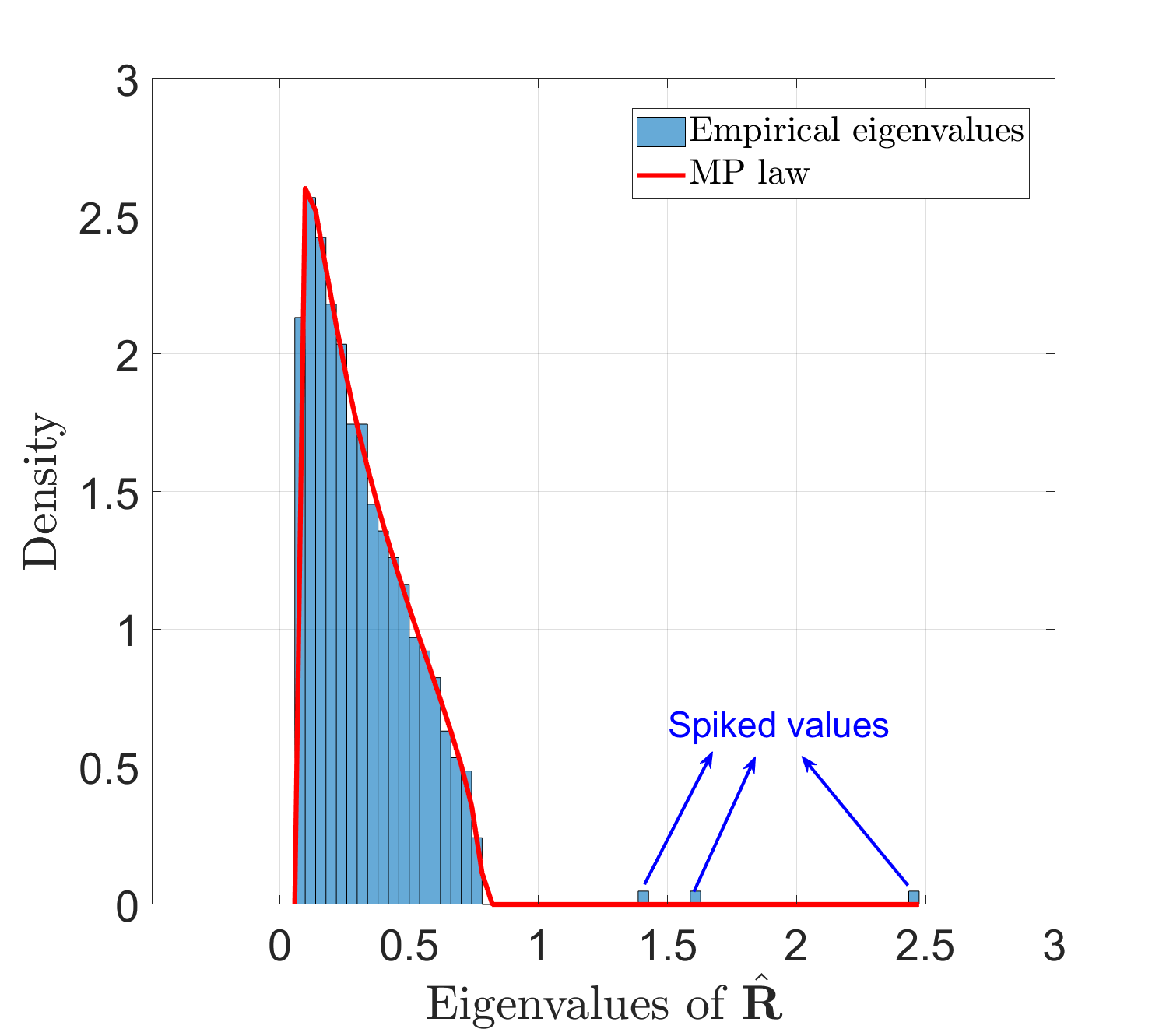}
\caption{Histogram of the eigenvalues of $\hat{\mathbf{R}}$.}
\label{fig_MP_law}
\end{figure}

It is well-known from~\cite{marchenko1967distribution} that for all $z\in\mathbb{C}\setminus\mathbb{R}$, 
\begin{align}\label{m(z)}
    \hat{m}_{\mathbf{R}}(z) \xrightarrow[N\rightarrow+\infty]{a.s.}  m(z), 
\end{align}
where $m(z)$ is the Stieltjes of a deterministic probability measure called the MP distribution, i.e.,
\begin{align}
    m(z) = \int_{\mathbb{R}}\frac{1}{\lambda-z}d\mu(\lambda), 
\end{align}
Here, the support of this distribution is the compact interval $[x^{-},x^{+}]$ with $x^{-}=\sigma_v^2(1-\sqrt{c})^2$ and $x^{+}=\sigma_v^2(1+\sqrt{c})^2$, which is defined by
\begin{align}
    &d\mu(x)\notag\\
    &\quad=\left(1-\frac{1}{c}\right)^{+}\delta_0 + \frac{\sqrt{(x-x^{-})(x^{+}-x)}}{2\sigma_v^2c\pi x}\pmb{1}_{[x^{-},x^{+}]}(x)dx, 
\end{align}
where $\left(1-\frac{1}{c}\right)^{+}\delta_0$ represents the cardinality of zero eigenvalues that can occur if $N-L>M$. Moreover, $m(z)$ satisfies the following fundamental equation
\begin{align}
    m(z) = \frac{1}{-z(1+\sigma_v^2cm(z))+\sigma_v^2(1-c)}. 
\end{align}
An equivalent statement of (\ref{m(z)}) is given with the following convergence in distribution
\begin{align}
    \hat{\mu}_{\mathbf{R}}(z) \xrightarrow[N\rightarrow+\infty]{\mathcal{D}}  \mu(z),
\end{align}
which holds almost surely, meaning that the empirical eigenvalue distribution of $\hat{\mathbf{R}}$ has the same asymptotic behavior as the MP distribution. Practically, for sufficiently large $M$ and $(N-L)$, the eigenvalue histogram of $\hat{\mathbf{R}}$ closely approximates the density of the MP distribution. This is illustrated in Fig.~\ref{fig_MP_law}, where we set $N=2048$, $M=N/4$, $L=N/4$, $K=5$, and  $\mathrm{SNR}=5$\,dB.  
\vspace{0.2cm}

\noindent{\bf Spectral behavior of the non-spiked eigenvalues}
    The MP distribution was originally derived as the limiting spectral distribution of the empirical eigenvalues of a pure noise covariance matrix. However, under Assumption~\ref{ass1}, where a deterministic perturbation of finite rank is introduced and this rank remains fixed as the matrix dimension increases, the MP law continues to describe the asymptotic distribution of the non-spiked eigenvalues. This well-established result can be rigorously justified using the Stieltjes transform of the spectral distribution. In the context of RMT, such models with finite-rank perturbations are commonly referred to as spiked models.

\vspace{0.2cm}
\noindent{\bf Localization of the spiked eigenvalues.}  
It is well known from the theory of spiked random covariance models (see, e.g.,~\cite{couillet2011random,yao2015sample,yang2018high}) that any population spike $t_i$ satisfying $t_i>\sqrt{c}$ generates, with probability tending to one as the dimensions grow, a spiked sample eigenvalue $\eta_i$ that appears as an \emph{outlier}, i.e., it separates from the bulk of the empirical spectral distribution. In particular, this sample outlier satisfies 
\[
\eta_i > \sigma_v^2(1+\sqrt{c})^{2}.
\]
Moreover, this outlier eigenvalue converges almost surely to the deterministic limit  
\begin{equation}
\label{eq:limit_eta}
\frac{\eta_i}{\sigma_v^2} \xrightarrow[N\to\infty]{a.s.} 1 + t_i + \frac{c(1+t_i)}{t_i}.
\end{equation}
The above relation is bijective in $t_i$, which makes it possible to construct a strongly consistent estimator of the population spike. Solving~\eqref{eq:limit_eta} for $t_i$ yields the following estimator:
\[
\hat{t}_i 
= \frac{\eta_i/\sigma_v^2 + 1 - c_M
+ \sqrt{\left(\eta_i/\sigma_v^2 + 1 - c_M\right)^2
- 4\,\eta_i/\sigma_v^2}}{2} - 1,
\]
Under the standard asymptotic regime, this estimator satisfies  
\[
|\hat{t}_i - t_i| \xrightarrow[N\to\infty]{a.s.} 0,
\]
thereby providing a consistent estimate of the population spike strength.

The purpose of this work is to provide a principled guideline for selecting the regularization parameter $\lambda$ in the semi-blind estimator of~\eqref{estimated channel vector}. A direct optimization of~\eqref{MSE_original} is not feasible, as it involves unobservable quantities such as the true channel matrix~$\mathbf{G}$ and the true noise subspace. To obtain a practically usable choice of $\lambda$, we exploit the asymptotic regime defined in Assumption~\ref{ass1} and characterize the limiting behavior of the MSE. The overall procedure is summarized as follows:

\begin{enumerate}
    \item Using Assumption~\ref{ass1}, we express the MSE in terms of its deterministic limit, which depends only on the population spike strengths $t_1,\ldots,t_K$ and their associated signal eigenvectors.
    \item We define the oracle regularization parameter $\bar{\lambda}^{\ast}$ as the minimizer of this deterministic limit.
    \item Finally, we construct a consistent estimator $\hat{\lambda}^{\ast}$ of $\bar{\lambda}^{\ast}$ by replacing the population quantities with their consistent sample-based estimates, namely the sample outlier eigenvalues $\eta_1,\ldots,\eta_K$ and the sample signal eigenvectors $\hat{\mathbf{u}}_{s1},\ldots,\hat{\mathbf{u}}_{sK}$.
\end{enumerate}

This yields a fully data-driven choice of the regularization parameter and, consequently, the channel estimate $\hat{\mathbf{G}}^{\ast}$.

\subsection{Deterministic Equivalent $\overline{MSE}(\lambda)$ and the Optimal $\bar{\lambda}^{\ast}$}
To derive the deterministic characterization $\overline{\mathrm{MSE}}(\lambda)$,We exploit the fact that, under Assumption~\ref{ass1}, the projection onto each sample signal eigenvector converges to its projection onto the corresponding population eigenvector. More precisely, for any deterministic vectors $\mathbf{a}$, the projection operator $\hat{\mathbf{u}}_{si}\hat{\mathbf{u}}_{si}^{H}$ satisfies
$$
\mathbf{a}^{H}\hat{\mathbf{u}}_{si}\hat{\mathbf{u}}_{si}^{H}\mathbf{a}
\;-\;
z_i\,\mathbf{a}^{H}{\mathbf{u}}_{si}{\mathbf{u}}_{si}^{H}\mathbf{a}
\;\xrightarrow[]{a.s.}\;0,
$$ 
where $z_i=\frac{1-c/t_i^2}{1+c/t_i}$.
Building on this projection convergence, we obtain the following asymptotic convergence result for the MSE:
\begin{theorem}\label{theorem_MSE}
[Deterministic equivalent] Let Assumption \ref{ass1_symbol} and \ref{ass1} hold. Then, we have 
\begin{align}
    \mathrm{sup}_{\lambda\in[0,1]} |\mathrm{MSE}(\lambda) - \overline{\mathrm{MSE}}(\lambda)|\xrightarrow[N\to\infty]{a.s.}{\longrightarrow} 0. 
\end{align}
where $\mathrm{sup}$ denotes the supremum, and 
\begin{align}\label{MSE_asy}
    \overline{\mathrm{MSE}}(\lambda) =& \frac{(1-\lambda)^2\sigma_v^2}{P_sK}\sum_{k=1}^K(1-z_k)t_k + \frac{\sigma_v^2\lambda^2\alpha}{a\beta},
\end{align}
with $z_k=\frac{1-c/t_k^2}{1+c/t_k}$. 
\begin{proof}
    See Appendix \ref{proof_MSE}. 
\end{proof}
\end{theorem}

To minimize (\ref{MSE_asy}) with respect to $\lambda$, we formulate the following optimization problem, i.e.,
\begin{align}
    \min_{\lambda} \,\, \overline{\mathrm{MSE}}(\lambda). 
\end{align}
Taking the derivative of $\overline{\mathrm{MSE}}(\lambda)$ with respect to $\lambda$ and setting it to zero yields
\begin{align}\label{derivative of MSE}
    \frac{\partial\overline{\mathrm{MSE}}(\lambda)}{\partial\lambda} =& \frac{2(\lambda-1)\sigma_v^2}{P_sK}\sum_{k=1}^K\left(1-z_k\right)t_k + \frac{2\sigma_v^2\lambda\alpha}{a\beta} \triangleq 0.
\end{align}
Solving this equation gives the optimal regularization parameter:
\begin{align}\label{asy_lambda}
    \bar{\lambda}^{\ast} = 1- \frac{\alpha\sigma_v^2}{\frac{a\beta\sigma_v^2}{P_sK}\sum_{k=1}^K(1-z_k)t_k+\alpha\sigma_v^2}.
\end{align}

\subsection{Estimated Optimal $\hat{\lambda}^{\ast}$ and Proposed Algorithm}
The optimal value $\bar{\lambda}^{\ast}$ in (\ref{asy_lambda}) depends on unobservable quantities, namely $z_k$ and $t_k$ for $k = 1, \ldots, K$, making it impractical for direct implementation. To overcome this limitation, we propose a consistent estimator that approximates the optimal $\bar{\lambda}^{\ast}$ using observable data. Specifically, the estimator is constructed based on the sample eigenvalues $\eta_i$ and the corresponding sample eigenvectors $\hat{\mathbf{u}}_{si}$ for $i = 1, \ldots, K$, enabling practical computation.

\begin{theorem}\label{theorem_lambda}
    [Estimated optimal regularization parameter] Under Assumption \ref{ass1_symbol} and \ref{ass1}, we have 
    \begin{align}
        |\hat{\lambda}^{\ast} - \bar{\lambda}^{\ast}| \overset{a.s.}{\longrightarrow} 0,
    \end{align}
where 
\begin{align}\label{lambda_estimate}
    \hat{\lambda}^{\ast} = 1- \frac{\sigma_v^2\alpha}{\frac{a\beta\sigma_v^2}{P_sK}\sum_{k=1}^K(1-\hat{z}_k)\hat{t}_k+\alpha\sigma_v^2}. 
\end{align}
with $\hat{z}_i=\frac{1-c/\hat{t}_i^2}{1+c/\hat{t}_i}$, and
\begin{align}\label{hat_t_i}
    \hat{t}_i = \frac{\eta_i/\sigma_v^2+1-c+\sqrt{(\eta_i/\sigma_v^2+1-c)^2-4\eta_i/\sigma_v^2}}{2} -1. 
\end{align}
\begin{proof}
    See Appendix~\ref{proof_lambda}. 
\end{proof}
\end{theorem}

\begin{theorem}\label{theorem_MSE_lambda}
    Under Assumption \ref{ass1_symbol} and \ref{ass1}, we have 
    \begin{align}\label{MSE_lambda}
        |\mathrm{MSE}(\hat{\lambda}^{\ast}) - \overline{\mathrm{MSE}}(\bar{\lambda}^{\ast})|\xrightarrow[N\to\infty]{a.s.} 0, 
    \end{align}
    where $\mathrm{MSE}(\hat{\lambda}^{\ast})$ is given by:
    \begin{align}
\mathrm{MSE}(\hat{\lambda}^{\ast}):= \frac{(1-\hat{\lambda}^{\ast})^2\sigma_v^2}{P_sK}\sum_{k=1}^K(1-\hat{z}_k)\hat{t}_k + \frac{\sigma_v^2(\hat{\lambda}^{\ast})^2\alpha}{a\beta}.\label{eq:MSE_lambda}
    \end{align}
    \begin{proof}
        See Appendix~\ref{proof_MSE_lambda}. 
    \end{proof}
\end{theorem}

Based on Theorems~\ref{theorem_MSE}, \ref{theorem_lambda}, and \ref{theorem_MSE_lambda}, we propose the following semi-blind estimator:
\begin{align}\label{optimal channel}
    \hat{\mathbf{G}}^{\ast} =& \left(\frac{1-\hat{\lambda}^{\ast}}{\hat{\lambda}^{\ast}}\hat{\mathbf{U}}_n\hat{\mathbf{U}}_n^H+\mathbf{I}_M\right)^{-1}\frac{1}{aL}\mathbf{Y}_p\mathbf{S}_p^H \notag\\
    =& \left(\hat{\lambda}^{\ast}\mathbf{I}_M+(1-\hat{\lambda}^{\ast})\hat{\mathbf{U}}_s\hat{\mathbf{U}}_s^H\right)\frac{1}{aL}\mathbf{Y}_p\mathbf{S}_p^H. 
\end{align}
According to Theorem \ref{theorem_MSE_lambda}, this estimator achieves, asymptotically, the minimum MSE among all semi-blind estimators of the form given in \eqref{estimated channel vector}.
Having established the key theoretical results underlying the proposed semi-blind estimator, we summarize the estimation steps in Algorithm~\ref{proposed algorithm}. A central advantage of this method is its simplicity: the resulting channel estimate is available in closed form and relies solely on simple operations involving the eigenvalues and eigenvectors of the sample covariance matrix.

\begin{algorithm}[t]
\caption{Proposed Channel Matrix Construction.} 
\label{proposed algorithm}
\begin{algorithmic}[1]
\STATE Compute the sample covariance matrix $\hat{\mathbf{R}}$ using the unknown data matrix $\mathbf{Y}_d$. 
\STATE Perform eigen-decomposition of $\hat{\mathbf{R}}$ to obtain the $i$th largest eigenvalue $\eta_i$ and its corresponding eigenvector $\hat{\mathbf{u}}_{si}$, for $i=1,\ldots,K$.
\STATE Compute $\hat{z}_i=\frac{1-c/\hat{t}_i^2}{1+c/\hat{t}_i}$ using 
\begin{align}
    \hat{t}_i = \frac{\eta_i/\sigma_v^2-1-c+\sqrt{(\eta_i/\sigma_v^2+1-c)^2-4\eta_i/\sigma_v^2}}{2}. \notag
\end{align}
\STATE Form the optimized regularization parameter $\hat{\lambda}^{\ast}$: 
\begin{align}
    \hat{\lambda}^{\ast} = 1- \frac{\sigma_v^2\alpha}{\frac{a\beta\sigma_v^2}{P_sK}\sum_{k=1}^K(1-\hat{z}_k)\hat{t}_k+\alpha\sigma_v^2}. \notag 
\end{align}
\STATE Construct the optimal estimate of the semi-blind channel matrix: 
\begin{align}
    \hat{\mathbf{G}}^{\ast} = \left(\hat{\lambda}^{\ast}\mathbf{I}_M+(1-\hat{\lambda}^{\ast})\hat{\mathbf{U}}_s\hat{\mathbf{U}}_s^H\right)\frac{1}{aL}\mathbf{Y}_p\mathbf{S}_p^H. 
\end{align}
\end{algorithmic}
\label{alg:1}
\end{algorithm}

\subsection{Scenario with Some $t_i$ Satisfying $0<t_i<\sqrt{c}$}
The previous results are derived under the assumption that, for large $M$, $t_i>\sqrt{c}$ for $i=1,\ldots, K$. However, in practical scenarios, it is generally unknown whether some of the spikes fall below this threshold, i.e., whether $t_i<\sqrt{c}$ for some $i$. In this subsection, we examine such scenarios and assess how the proposed algorithm performs in the presence of weak spikes. Specifically, suppose there exists $K_2<K$ such that, for sufficiently large $M$, 
\begin{align}
    t_1>t_2>\cdots>t_{K-K_2}>\sqrt{c}>t_{K-K_2+1}>\cdots>t_K>0.  
\end{align}
In this regime, the population eigenvectors $\mathbf{u}_{si}$ associated with the weak spikes (i.e., those with $t_i<\sqrt{c}$) no longer asymptotically align with their sample counterparts $\hat{\mathbf{u}}_{si}$. Similarly, the sample eigenvalues $\lambda_i$ are no longer informative about the corresponding $t_i$, for $i=K-K_2+1,\ldots, K$. In particular, the work in ~\cite{couillet2011random} shows that: 
\begin{align}\label{u_hat_u}  \mathbf{u}_{si}^H\hat{\mathbf{u}}_{si}\hat{\mathbf{u}}_{si}^H\mathbf{u}_{si} \overset{a.s.}{\longrightarrow} 0, \,\, i=K-K_2+1,\ldots,K,
\end{align}
while 
\begin{align}\label{lambda_c}
    \frac{\eta_i}{\sigma_v^2} \overset{a.s.}{\longrightarrow} (1+\sqrt{c})^2, \ \ i=K-K_2+1,\ldots K,
\end{align}
rather than obeying (\ref{lambda_t_i}). 

The population covariance matrix is written as 
\begin{align}
    \mathbf{R} = \sigma_v^2\Big(\sum_{i=1}^{K-K_2}t_i\mathbf{u}_{si}\mathbf{u}_{si}^H+\sum_{j=K-K_2+1}^Kt_j\mathbf{u}_{sj}\mathbf{u}_{sj}^H\Big). 
\end{align}
Following similar steps as in the proof of Theorem~\ref{theorem_MSE}, and incorporating the asymptotic behaviors in (\ref{u_hat_u}) and (\ref{lambda_c}), the deterministic equivalent of $\mathrm{MSE}(\lambda)$ becomes:
\begin{align}
    \overline{\mathrm{MSE}}(\lambda) =& \frac{(1-\lambda)^2\sigma_v^2}{P_sK}\sum_{k=1}^Kt_k + \frac{\lambda^2\sigma_v^2\alpha}{a\beta} \notag\\
    &-\frac{(1-\lambda)^2\sigma_v^2}{P_sK}\sum_{k=1}^{K-K_2}z_kt_k, \label{eq:MSE_non_spike}
\end{align}
where $z_k$ and $t_k$ retain the same definitions as in Theorem~\ref{theorem_MSE}, but now the summation limits are adjusted to reflect the effective number of strong spikes $(K-K_2)$. 
\begin{align}
    \frac{1}{K}\sum_{k=1}^Kt_k = \frac{P_s}{\sigma_v^2K}\sum_{k=1}^K\mathbf{g}_k^H\mathbf{g}_k, 
\end{align}
In this case, we cannot rely on the estimation of $t_k$ from $\eta_k$ to approximate 
$\frac{P_s}{\sigma_v^2 K}\sum_{k=1}^K\mathbf{g}_k^H\mathbf{g}_k$. 
Indeed, for all $k \geq K-K_2+1$, the corresponding $\eta_k$ converge to the right edge of the bulk,
\begin{align}
    \sigma_v^2(1+\sqrt{c})^2,
\end{align}
a limit that is completely independent of the underlying values $t_k$.  
As a result, the eigenvalue-based method becomes unreliable for estimating $\frac{1}{K}\sum_{k=1}^{K}{\bf g}_k^{H}{\bf g}_k$.

We therefore turn to the training-based estimator, which, as we show below, remains sufficient to construct a consistent estimate of 
$\frac{1}{K}\sum_{k=1}^{K}{\bf g}_k^{H}{\bf g}_k$.

From~\eqref{G_training} and the identity ${\bf S}_p{\bf S}_p^{H}=aL{\bf I}_K$, the training-based channel estimate satisfies
\[
\frac{1}{aL}{\bf S}_p{\bf Y}_p^{H}
  ={\bf G}^{H}+\frac{{\bf S}_p{\bf V}_p^{H}}{aL}.
\]
Consequently,
\[
\frac{1}{a^2L^2}{\bf S}_p{\bf Y}_p^{H}{\bf Y}_p{\bf S}_p^H
= \left({\bf G}^{H}+\frac{{\bf S}_p{\bf V}_p^{H}}{aL}\right)
  \left({\bf G}+\frac{{\bf V}_p{\bf S}_p^{H}}{aL}\right).
\]

Using standard concentration results for bilinear forms of random vectors \cite{couillet2011random}, we have
\[
\frac{{\bf S}_p{\bf V}_p^{H}}{aL}\,{\bf G} \xrightarrow[N\to\infty]{a.s.} {\bf 0},
\qquad
{\bf G}^{H}\frac{{\bf V}_p{\bf S}_p^{H}}{aL} \xrightarrow[N\to\infty]{a.s.} {\bf 0},
\]
and
\[
\frac{{\bf S}_p{\bf V}_p^{H}{\bf V}_p{\bf S}_p^{H}}{(aL)^2}
- \frac{\sigma_v^2\alpha}{a\beta}\mathbf{I}_K
\xrightarrow[N\to\infty]{a.s.} {\bf 0}.
\]

Therefore, a consistent estimator of $\frac{1}{K}\operatorname{tr}({\bf G}^{H}{\bf G})$ is
\[
\frac{1}{a^2L^2}\operatorname{tr}\!\left({\bf S}_p{\bf Y}_p^{H}{\bf Y}_p{\bf S}_p^{H}\right)
  - \frac{\sigma_v^2\alpha}{a\beta}.
\]
Substituting this expression into \eqref{eq:MSE_non_spike}, we obtain the following consistent estimator for the MSE:
\begin{align}
\widehat{\rm MSE}(\lambda):=&(1-\lambda)^2\left(\frac{1}{a^2L^2}{\rm tr}({\bf S}_p{\bf Y}_p^{H}{\bf Y}_p{\bf S}_p^{H})-\frac{\sigma_v^2\alpha}{a\beta}\right) \nonumber\\
&+\frac{\lambda^2\sigma_v^2\alpha}{a\beta}-\frac{(1-\lambda)^2\sigma_v^2}{P_sK}\sum_{k=1}^{K-K_2}\hat{z}_k\hat{t}_k.
\end{align}
where in the last term $\hat{z}_k$ and $\hat{t}_k, k=1,\ldots, K-K_2$ is justified because these indices correspond to spikes lying outside the bulk, for which consistent estimation is possible. Setting $\lambda$ to the value that minimizes $\widehat{\mathrm{MSE}}(\lambda)$ yields
\begin{align}
    &\hat{\lambda}^{\ast} 
    = 1\nonumber \\
    &- \frac{\sigma_v^2 \alpha}{
    \beta\Big( \frac{1}{aL^2}\mathrm{tr}({\bf S}_p\mathbf{Y}_p^H\mathbf{Y}_p{\bf S}_p^{H})
    + \frac{\sigma_v^2\alpha}{\beta} \Big)
    - \frac{a\beta\sigma_v^2}{P_s K}\sum\limits_{i=1}^{K-K_2}\hat{z}_i\hat{t}_i
    + \sigma_v^2\alpha }.\label{eq:lambda_optimal}
\end{align}
We then define the following semi-blind channel estimator:
\begin{align}
    \hat{\mathbf{G}}^{\ast} 
    = \left( \hat{\lambda}^{\ast}\mathbf{I}_M 
    + (1-\hat{\lambda}^{\ast})\hat{\mathbf{U}}_s\hat{\mathbf{U}}_s^H \right)
    \frac{1}{aL}\mathbf{Y}_p\mathbf{S}_p^H.
\end{align}
Finally, we note that the quantity $K-K_2$, corresponding to the number of spikes satisfying 
$t_i \geq \sqrt{c}$, can be directly inferred from the empirical eigenvalue distribution of 
the sample covariance matrix. In practice, $K_2$ is simply the number of sample eigenvalues 
that lie above the right edge of the bulk, namely $\sigma_v^2(1+\sqrt{c})^2$. Equivalently, 
the remaining $K_2$ eigenvalues are absorbed into the bulk and thus correspond to the 
non-detectable spikes. This provides a simple and robust criterion for identifying the 
detectable components, and Algorithm~\ref{alg:1} can be easily adapted to incorporate this 
detection step. 

\begin{algorithm}[t]
\caption{Proposed Channel Matrix Construction.} 
\label{proposed algorithm}
\begin{algorithmic}[1]
\STATE Compute the sample covariance matrix $\hat{\mathbf{R}}$ using the unknown data matrix $\mathbf{Y}_d$. 
\STATE Perform eigen-decomposition of $\hat{\mathbf{R}}$ to obtain the $i$th largest eigenvalue $\eta_i$ and its corresponding eigenvector $\hat{\mathbf{u}}_{si}$, for $i=1,\ldots,K$.
\STATE Estimate $K_1:=K-K_2$ the number of spikes by $\hat{K}_1:=\#\{\eta_i> \sigma_v^2(1+\sqrt{c})^2\}$.
\STATE Compute $\hat{z}_i=\frac{1-c/\hat{t}_i^2}{1+c/\hat{t}_i}, i=1,\cdots K_2$ using 
\begin{align}
    \hat{t}_i = \frac{\eta_i/\sigma_v^2-1-c+\sqrt{(\eta_i/\sigma_v^2+1-c)^2-4\eta_i/\sigma_v^2}}{2}.\notag
\end{align}
\STATE Compute $\hat{\lambda}^{\ast}$ using the expression in \eqref{eq:lambda_optimal}.
\STATE Construct the optimal estimate of the semi-blind channel matrix: 
\begin{align}
    \hat{\mathbf{G}}^{\ast} = \left(\hat{\lambda}^{\ast}\mathbf{I}_M+(1-\hat{\lambda}^{\ast})\hat{\mathbf{U}}_s\hat{\mathbf{U}}_s^H\right)\frac{1}{aL}\mathbf{Y}_p\mathbf{S}_p^H. 
\end{align}
\end{algorithmic}
\label{alg:1}
\end{algorithm}

\subsection{Complexity Analysis}
In this subsection, we analyze the computational complexity of the two proposed channel matrix construction algorithms, namely Algorithms~1 and~2. The complexity is derived by separately evaluating the major computational components, including sample covariance matrix construction, dominant signal subspace extraction, pilot-based channel estimation, regularization parameter computation, and semi-blind channel reconstruction. Compared with Algorithm~1, Algorithm~2 additionally performs spike detection and evaluates an extra trace term for regularization parameter estimation.

The sample covariance matrix $\hat{\mathbf{R}}=\frac{1}{N-L}\mathbf{Y}_d\mathbf{Y}_d^{H}$ is first constructed, requiring $\mathcal{O}(M^{2}(N-L))$ operations. Since only the $K$ dominant eigenpairs are required by the proposed algorithms, performing a full eigendecomposition is unnecessary. Instead, a partial eigendecomposition algorithm, such as the power method or the Lanczos method, can be employed to extract the dominant signal subspace. According to~\cite{xu1994fast}, the corresponding computational complexity is approximately $\mathcal{O}(M^{2}K)$.

The pilot-based channel estimate $\hat{\mathbf{G}}_{\mathrm{training}}=\frac{1}{aL}\mathbf{Y}_{p}\mathbf{S}_{p}^{H}$ is obtained by multiplying an $M\times L$ matrix with an $L\times K$ matrix, resulting in a computational complexity of $\mathcal{O}(MLK)$. The estimated optimal regularization parameter $\hat{\lambda}^{\ast}$ is then computed using the extracted dominant eigenvalues together with the pilot-based statistics. For Algorithm~1, the associated scalar operations on the $K$ dominant eigenvalues require only $\mathcal{O}(K)$ operations. For Algorithm~2, the regularization parameter additionally involves the trace term $\mathrm{tr}(\mathbf{S}_p\mathbf{Y}_p^{H}\mathbf{Y}_p\mathbf{S}_p^{H})=\|\mathbf{Y}_p\mathbf{S}_p^{H}\|_F^2$. Since the matrix product $\mathbf{Y}_{p}\mathbf{S}_{p}^{H}$ has already been computed during pilot-based channel estimation, evaluating this trace term requires only an additional $\mathcal{O}(MK)$ operations. This cost is dominated by the $\mathcal{O}(MLK)$ complexity of the pilot-based channel estimation and therefore does not change the overall computational order.

Finally, the semi-blind channel estimate is constructed as $\hat{\mathbf{G}}=\hat{\lambda}^{\ast}\hat{\mathbf{G}}_{\mathrm{training}}+(1-\hat{\lambda}^{\ast})\hat{\mathbf{U}}_{s}\hat{\mathbf{U}}_{s}^{H}\hat{\mathbf{G}}_{\mathrm{training}}$. Rather than explicitly forming the $M\times M$ projection matrix $\hat{\mathbf{U}}_{s}\hat{\mathbf{U}}_{s}^{H}$, the projection is efficiently evaluated as $\hat{\mathbf{U}}_{s}(\hat{\mathbf{U}}_{s}^{H}\hat{\mathbf{G}}_{\mathrm{training}})$, which requires two matrix multiplications. Since $\hat{\mathbf{U}}_{s}\in\mathbb{C}^{M\times K}$ and $\hat{\mathbf{G}}_{\mathrm{training}}\in\mathbb{C}^{M\times K}$, the corresponding computational complexity is $\mathcal{O}(MK^{2})$. The overall computational complexity of both Algorithm~1 and Algorithm~2 is therefore $\mathcal{O}(M^{2}(N-L)+M^{2}K+MLK+MK^{2})$.

\par

\begin{figure}
\centering
\includegraphics[width=1.0\linewidth]{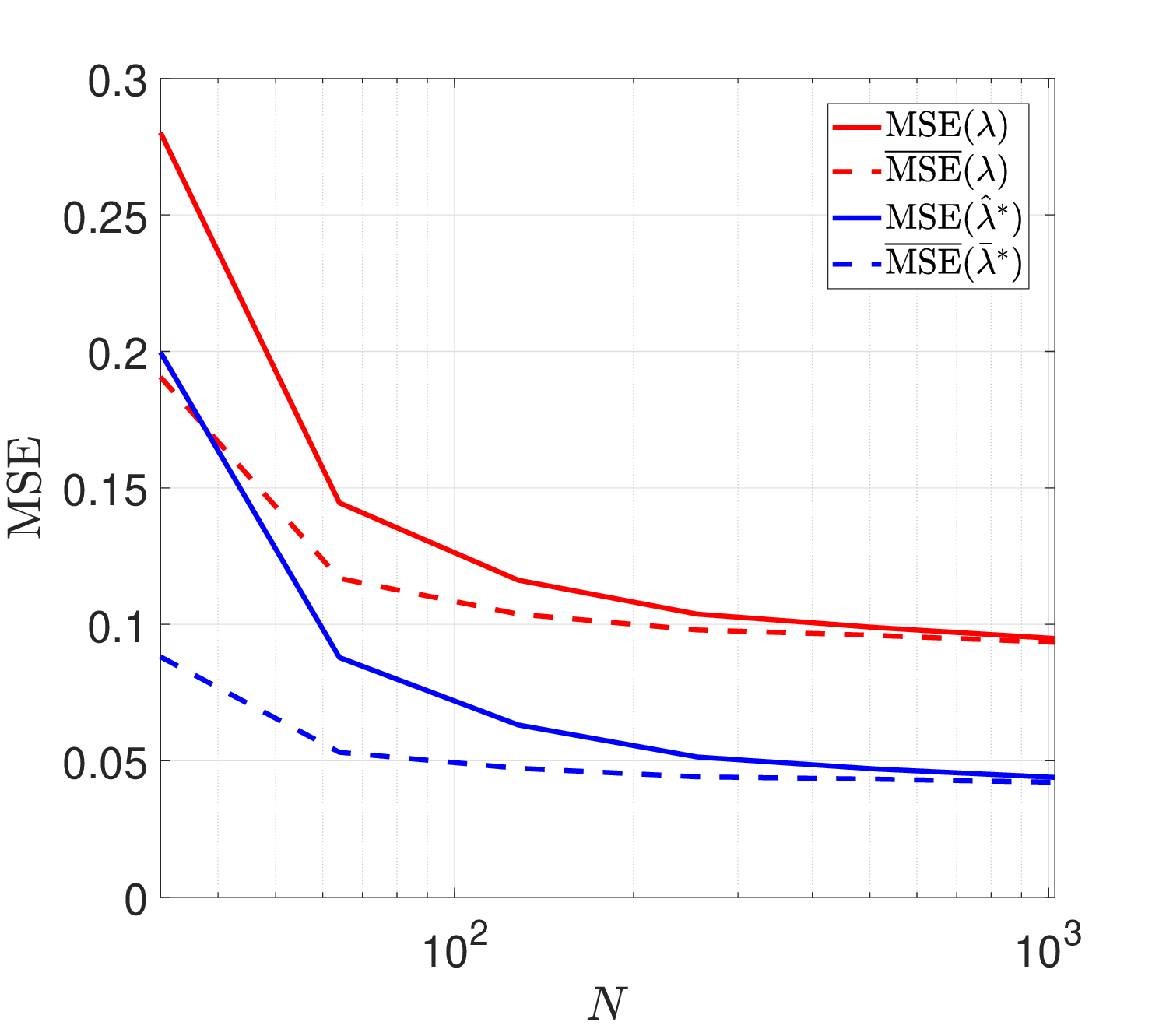}
\caption{$\mathrm{MSE}$-performance versus $N$ when $K=3$, $\alpha=1/2$, $\beta=1/8$, and $\mathrm{SNR}=15$\,dB.}
\label{fig_N_MSE}
\end{figure}

\begin{figure}
\centering
\includegraphics[width=1.0\linewidth]{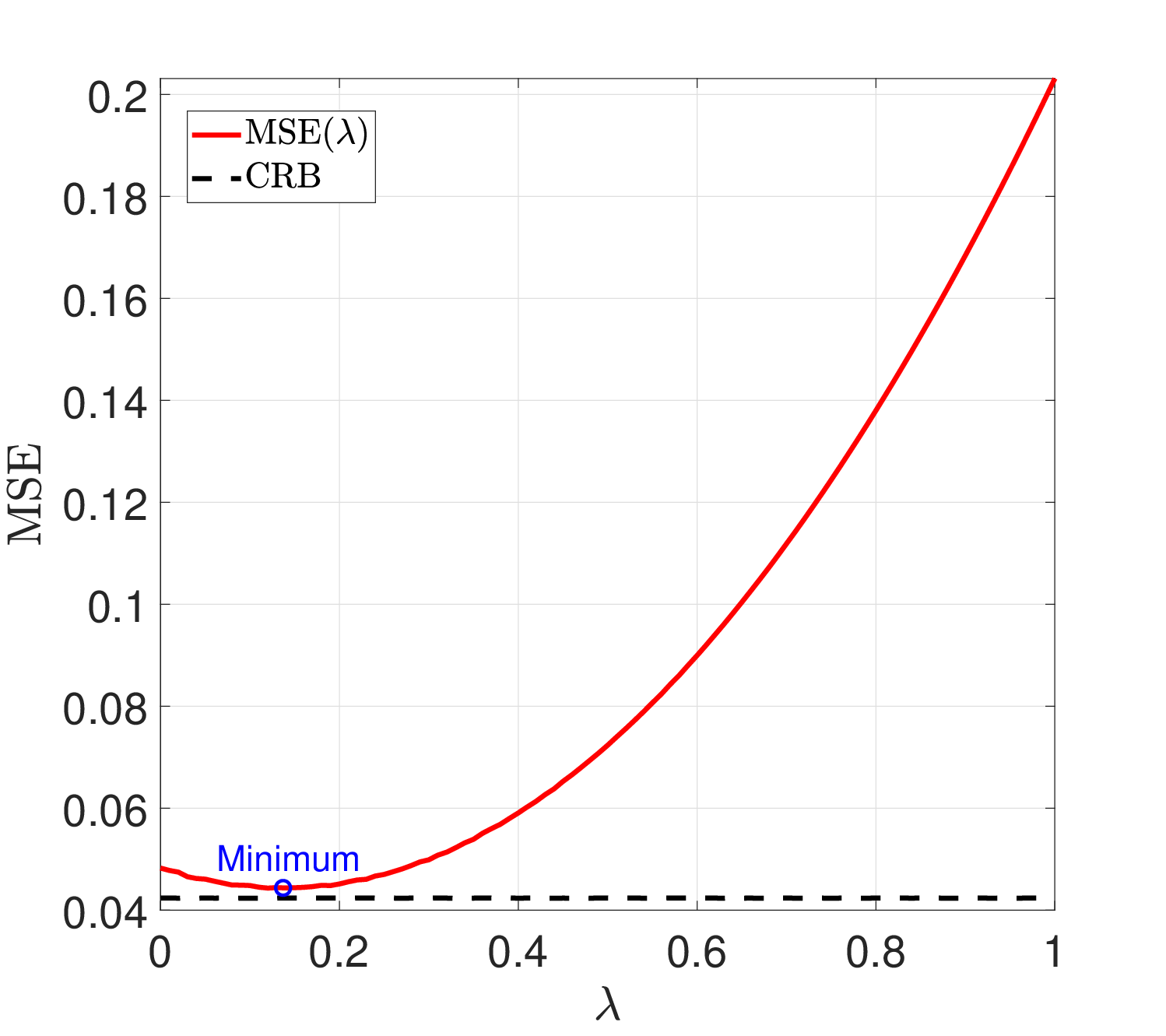}
\caption{$\mathrm{MSE}$-performance versus $\lambda$ when $K=3$, $\alpha=1/4$, $\beta=1/8$, $N=256$, and $\mathrm{SNR}=15$\,dB.}
\label{fig_lambda_MSE}
\end{figure}

\section{Simulation results}
\label{sec_sim}
This section presents numerical simulations to evaluate the asymptotic behavior and finite-dimensional performance of the proposed semi-blind channel estimation algorithm under representative NTN propagation conditions. The simulations are conducted using the open-source Sionna/OpenNTN framework, which implements the standardized NTN channel models specified in 3GPP TR~38.811~\cite{3gpp2020study,due2025open}. Unless otherwise specified, the Dense Urban (DUR) propagation scenario is considered for uplink transmission. The carrier frequency is set to $2$~GHz, the satellite BS is located at an altitude of $600$~km, and the user elevation angle is fixed at $10^\circ$. For the considered satellite altitude and elevation angle, the corresponding satellite--user slant range is approximately $1930$~km. The satellite BS employs a vertically polarized $1\times512$ aperture antenna array, whereas each user terminal is equipped with a single vertically polarized omnidirectional antenna. According to the selected propagation scenario and satellite--user geometry, the adopted NTN channel model generates the corresponding propagation characteristics following the statistical procedures specified in 3GPP TR~38.811. These characteristics include the LoS/NLoS condition, path loss, shadow fading, delay spread, path delays, cluster powers, angular spreads, and Doppler effects. For each channel realization, the resulting channel response is mapped to the narrowband signal model considered in this paper, yielding the channel matrix $\mathbf{G}$ used for channel estimation. Unless otherwise stated, the theoretical and numerical evaluations adopt the block-fading model, under which $\mathbf{G}$ is assumed to remain approximately constant over the pilot and data observations within one effective coherence block. The impact of channel variation within the observation interval is separately investigated in Section~V-C. The pilot matrix $\mathbf{S}_p$ is chosen to be orthogonal with a power scaling factor $a=1$, while the data symbols $\mathbf{S}_d$ with $P_s=1$ are drawn from a quadrature phase-shift keying (QPSK) constellation. The signal-to-noise ratio (SNR) is defined as $\mathrm{SNR} = \frac{(a\beta+P_s(1-\beta))\|\mathbf{G}\|_F^2}{\sigma_v^2}$. Each simulation point is obtained by averaging over $1000$ independent Monte Carlo realizations.

\begin{figure}[t]
\centering
\includegraphics[width=1.0\linewidth]{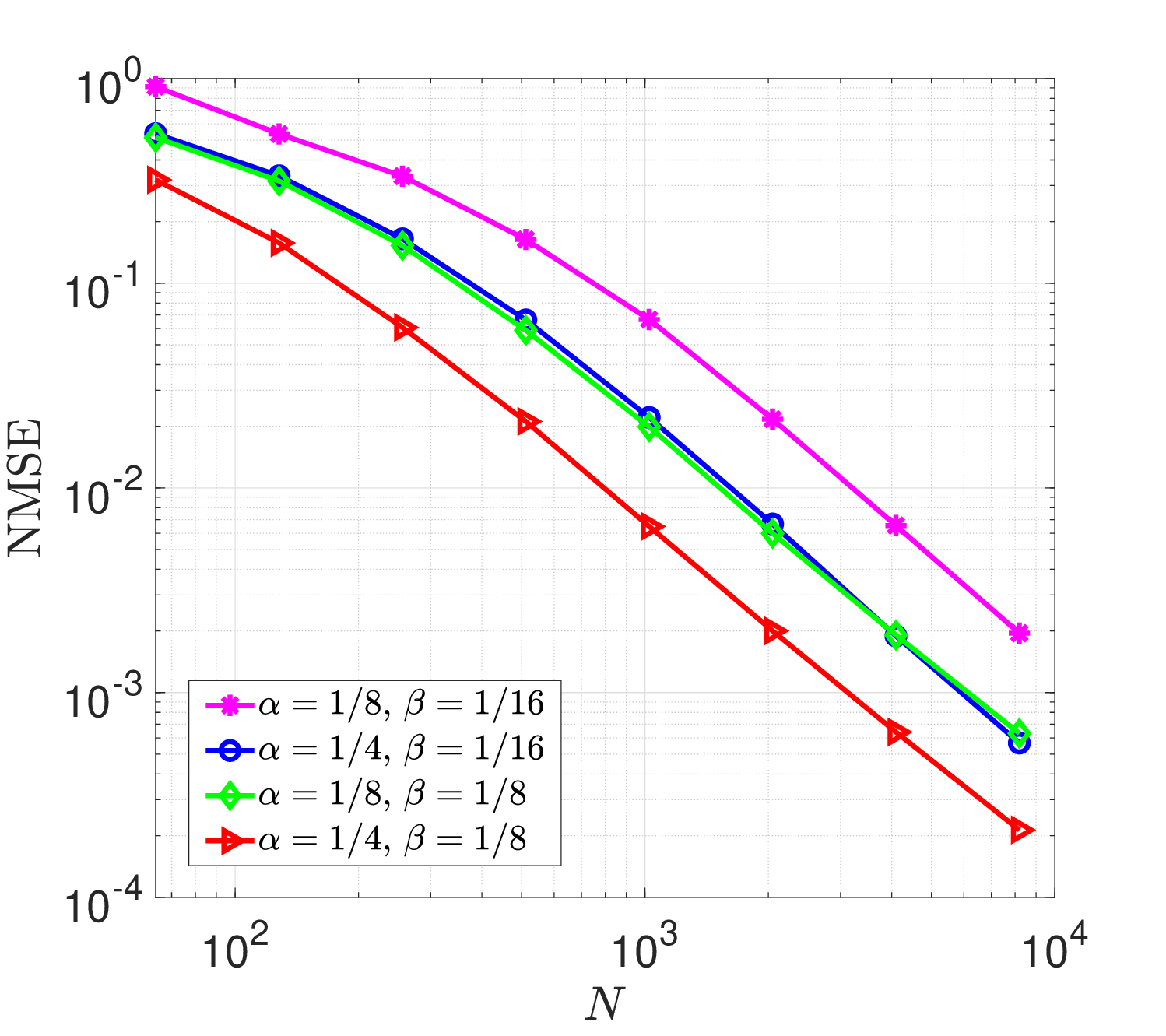}
\caption{$\mathrm{NMAE}$ versus $N$ under different $\alpha$ and $\beta$ with $K=3$ and $\mathrm{SNR}=15$\,dB.}
\label{fig_N_NMAE}
\end{figure}

\begin{figure}[t]
\centering
\includegraphics[width=1.0\linewidth]{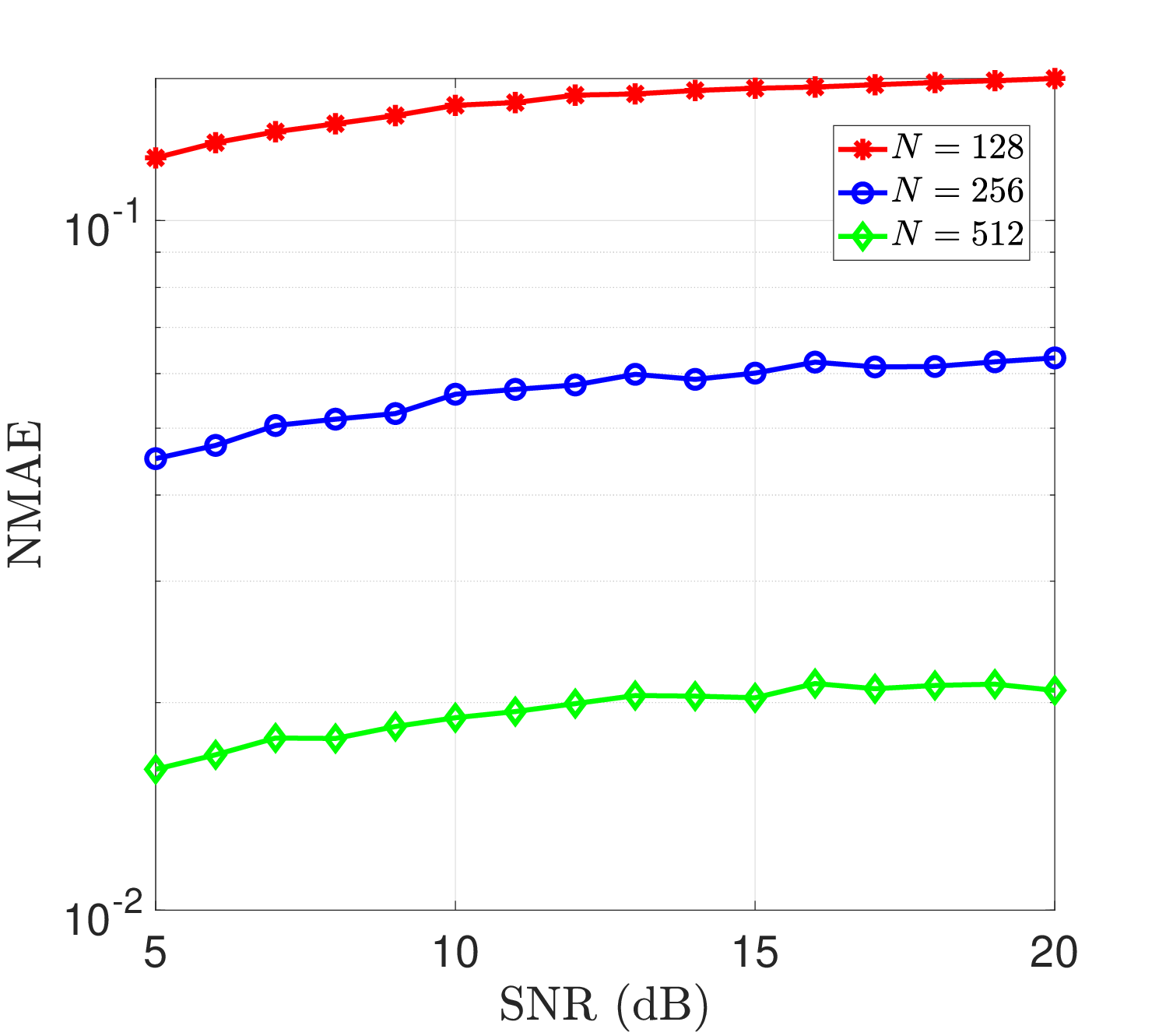}
\caption{$\mathrm{NMAE}$ versus $\mathrm{SNR}$ under different $N$ with $K=3$, $\alpha=1/4$, and $\beta=1/8$.}
\label{fig_SNR_NMAE}
\end{figure}

\begin{figure}[t]
\centering
\includegraphics[width=1.0\linewidth]{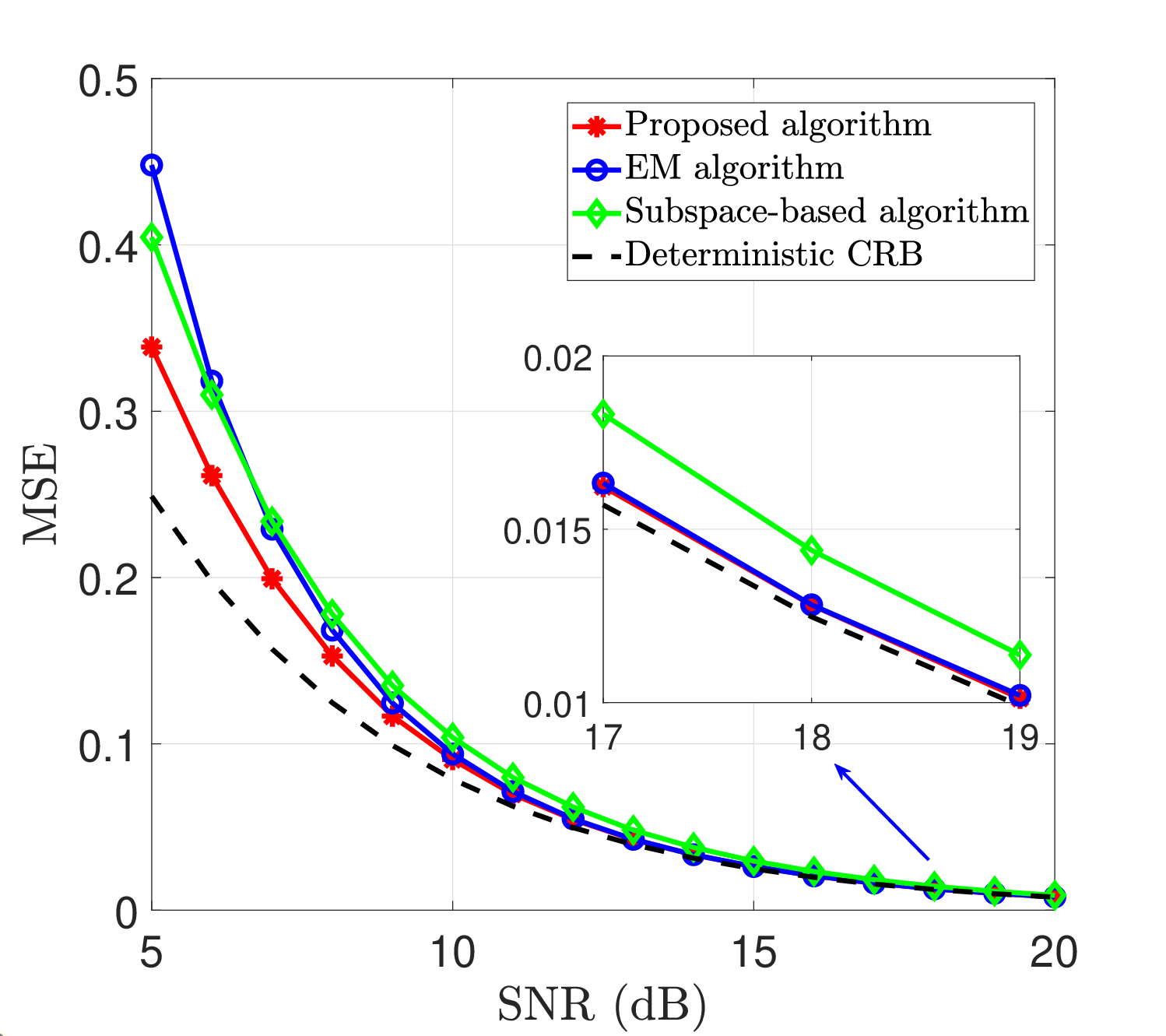}
\caption{$\mathrm{MSE}$-performance versus $\mathrm{SNR}$ when $K=3$, $\alpha=1/4$, $\beta=1/8$, and $N=512$.}
\label{fig_SNR_MSE}
\end{figure}

\subsection{Experiments for Validating Accuracy}
In the first experiment, we investigate the asymptotic behavior of $\mathrm{MSE}(\hat{\lambda})$. Under the asymptotic regime specified by Assumption~\ref{ass1}, the system parameters are set as $\mathrm{SNR}=15$\,dB, $\alpha = M/N = 1/2$, $\beta = L/N = 1/8$, and $c = M/(N-L) = 4/7$. The number of users is fixed at $K = 3$. As shown in Fig.~\ref{fig_N_MSE}, we compare the results predicted by Theorem~\ref{theorem_MSE} and Theorem~\ref{theorem_lambda} under both a randomly selected regularization parameter $\lambda$ and the optimal regularization parameter $\lambda^\ast$. The random value of $\lambda$ is selected once prior to the Monte Carlo simulations and kept fixed throughout all Monte Carlo iterations. It is observed that as the total transmission block length $N$ increases, the MSEs of the proposed algorithm converge to their respective asymptotic limits. Furthermore, Fig.~\ref{fig_lambda_MSE} shows the variation of the MSE as a function of $\lambda$, ranging from $0$ to $1$ in increments of $0.01$, and compares the results with the deterministic Cramér-Rao bound (CRB) derived  in\cite{zhang2025fundamental}. Here, $K=3$, $N=256$, $\alpha=1/2$, $\beta=1/4$, and $\mathrm{SNR}=15$\,dB. The minimum MSE is achieved at $\lambda = 0.13$, which is in close agreement with the theoretical optimum $\hat{\lambda}^\ast = 0.1350$ derived from~(\ref{asy_lambda}). The corresponding minimum MSE closely approaches the deterministic CRB, demonstrating the high estimation accuracy of the proposed algorithm.

Next, we evaluate the accuracy of the asymptotic approximation to the MSE by computing the normalized MSE approximation error (NMAE), defined as:
\begin{align}
\mathrm{NMAE} = \frac{\mathbb{E}\big[|\mathrm{MSE}(\lambda)-\overline{\mathrm{MSE}}(\lambda)|^2\big]}{\mathbb{E}\big[\mathrm{MSE}(\lambda)^2\big]}.
\end{align}
As shown in Fig.~\ref{fig_N_NMAE} with $K=3$ and $\mathrm{SNR}=15$\,dB, the NMAE decreases as the system dimensions increase, confirming the validity and accuracy of the asymptotic expression across various values of $\alpha$ and $\beta$. In addition, Fig.~\ref{fig_SNR_NMAE} presents the variation of the NMAE with respect to SNR, ranging from 5 dB to 20 dB with an increment of $1$\,dB, under different values of $N$. The results demonstrate that the asymptotic approximation remains accurate over a broad range of SNRs.

\begin{figure}[t]
\centering
\includegraphics[width=1.0\linewidth]{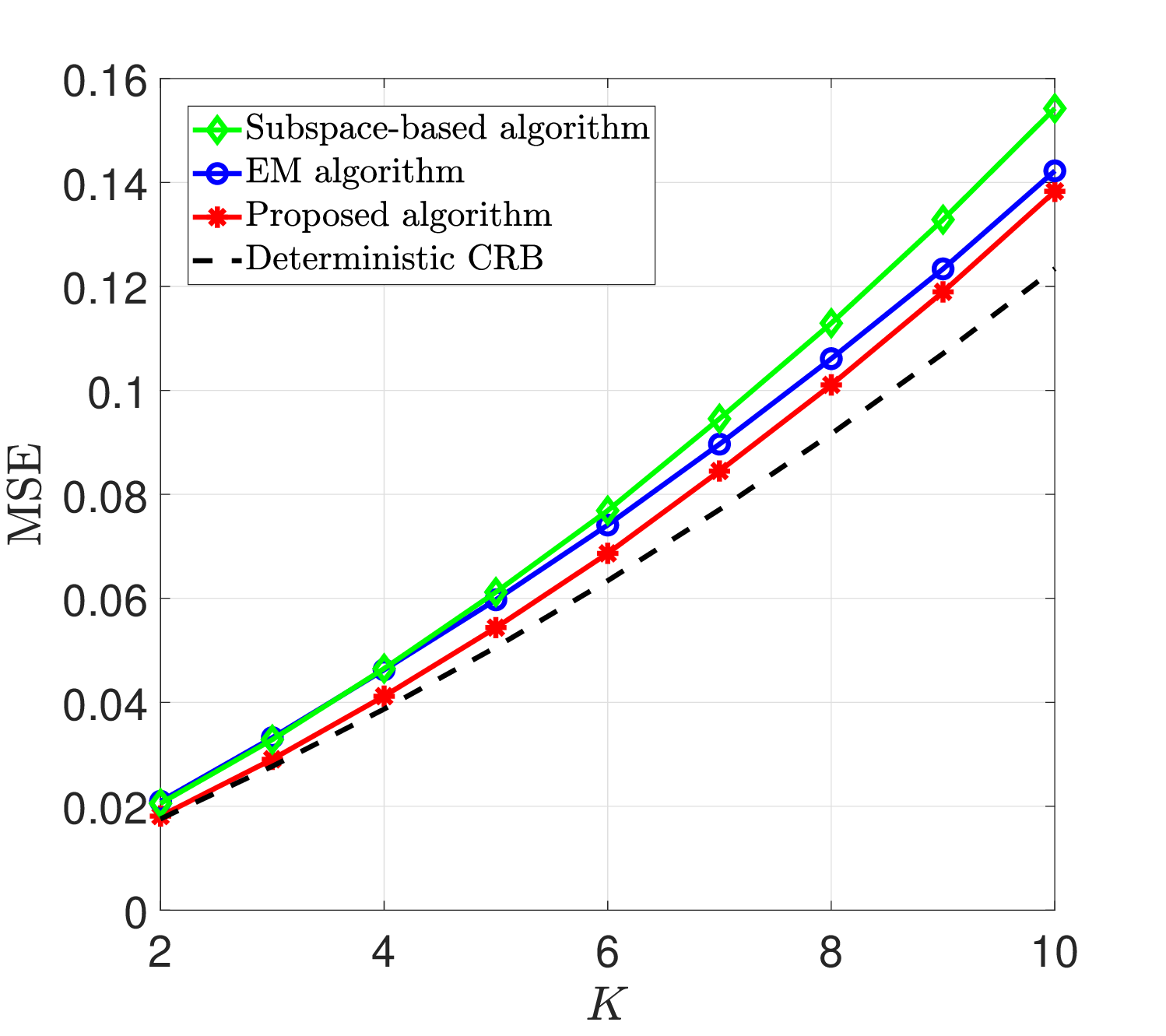}
\caption{$\mathrm{MSE}$-performance versus $K$ when $\alpha=1/4$, $\beta=1/8$, $N=1024$, and $\mathrm{SNR}=15$\,dB.}
\label{fig_K_MSE}
\end{figure}

\subsection{Comparison with Baseline Methods}
In this subsection, we compare the MSE performance of the proposed algorithm with that of the EM algorithm presented in~\cite{nayebi2017semi} and the subspace-based algorithm proposed in~\cite{zhong2024subspace}. Fig.~\ref{fig_SNR_MSE} illustrates the MSE results for SNR values ranging from $5$\,dB to $20$\,dB, in increments of $1$\,dB. The simulation parameters are set to $N = 512$, $K = 3$, $\alpha = 1/4$, and $\beta = 1/8$. As the SNR increases, all algorithms show improved estimation accuracy. Notably, the proposed algorithm consistently achieves lower MSE across the entire SNR range, demonstrating its superior performance. Additionally, Fig.~\ref{fig_K_MSE} depicts the MSE performance as the number of users $K$ increases from $2$ to $10$ with a step size of $1$, with the SNR fixed at $15$ dB. It can be observed that the MSE of all three algorithms decreases as $K$ increases. Again, the proposed algorithm consistently outperforms both the EM and subspace-based algorithms across all tested values of $K$. 


\subsection{Impact of Pilot Overhead and Channel Temporal Variation}
We next investigate the impact of channel temporal variation and pilot overhead on the proposed semi-blind estimator. To decouple the effect of pilot overhead from the number of data samples used for covariance estimation, the data interval is fixed at $N_d=128$, while the pilot ratio $\beta=L/N=L/(L+N_d)$ is varied by adjusting the pilot length $L$. The remaining parameters are set to $M=128$, $K=3$, and $\mathrm{SNR}=10$~dB. To emulate different effective channel coherence conditions, the channel during the data interval is modeled according to the first-order temporally correlated process $\mathbf{G}[n]=\rho\mathbf{G}[n-1]+\sqrt{1-\rho^2}\mathbf{E}[n]$, where $\mathbf{E}[n]$ denotes an independent channel innovation with the same second-order statistics as $\mathbf{G}[n]$, and $\rho\in[0,1]$ denotes the temporal correlation coefficient. Three correlation levels are considered to represent low, moderate, and high temporal variations, respectively. A smaller $\rho$ corresponds to a faster decay of the temporal channel correlation and hence a shorter effective coherence interval.

\begin{figure}[t]
\centering
\includegraphics[width=1.0\linewidth]{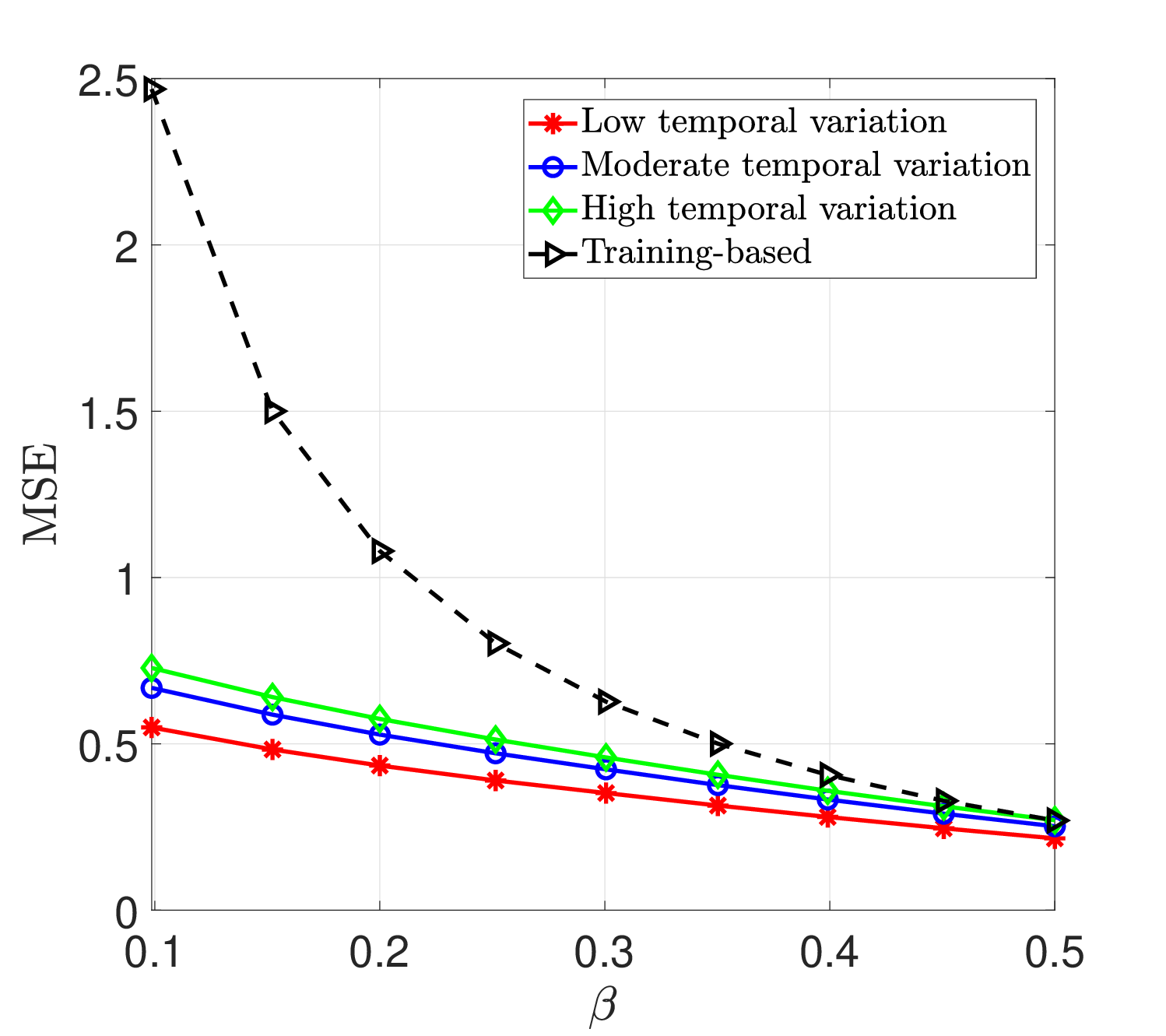}
\caption{$\mathrm{MSE}$-performance versus $\beta$ under different channel temporal variations when $K=3$, $M=128$, and $\mathrm{SNR}=10$\,dB.}
\label{fig_beta_MSE2}
\end{figure}

\begin{figure}[t]
\centering
\includegraphics[width=1.0\linewidth]{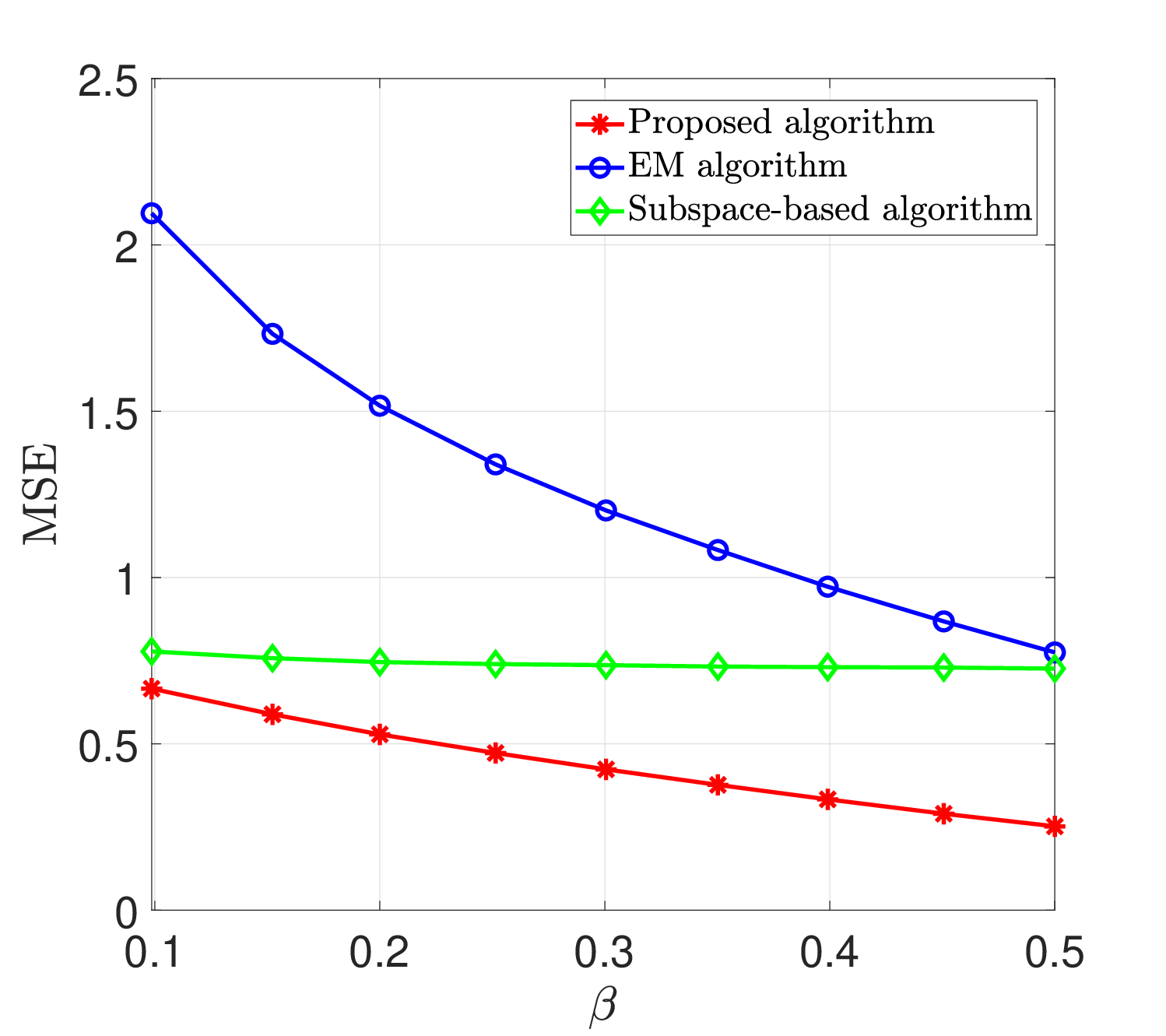}
\caption{$\mathrm{MSE}$-performance versus $\beta$ under moderate channel temporal variation when $K=3$, $M=128$, and $\mathrm{SNR}=10$\,dB.}
\label{fig_beta_MSE}
\end{figure}

Fig.~\ref{fig_beta_MSE} shows the MSE of the proposed estimator versus the pilot ratio under different temporal variation levels, together with the pilot-based estimator. As the temporal variation becomes faster, the MSE increases because the received data samples become less temporally consistent with the channel observed during the pilot interval, thereby reducing the reliability of the data-aided covariance information. The performance degradation is more pronounced at small pilot ratios, where the proposed estimator relies more strongly on the covariance information extracted from the received data. As $\beta$ increases, the pilot-based channel estimate becomes more accurate, and the performance differences among the considered temporal variation levels gradually decrease. These results show that the proposed estimator benefits most from data-aided covariance information under relatively slow channel variations, whereas increasing the pilot overhead improves robustness as the effective coherence interval becomes shorter.

Fig.~\ref{fig_beta_MSE2} further compares the proposed estimator with the EM and subspace-based algorithms under the moderate temporal-variation condition. The MSE of all considered methods decreases as the pilot ratio increases due to the improved pilot-based channel estimate. The proposed estimator consistently achieves the lowest MSE over the considered range of pilot ratios, with a more pronounced advantage in the low-pilot-overhead regime. This gain results from the additional covariance information extracted from the received data, which complements the limited pilot observations. As the pilot ratio increases, all considered methods benefit from the additional pilot observations, while the proposed estimator maintains its performance advantage.

Taken together, Figs.~\ref{fig_beta_MSE} and~\ref{fig_beta_MSE2} demonstrate the tradeoff between pilot overhead and channel temporal variation. Under relatively slow channel variations, the proposed estimator can exploit the data-aided covariance information more effectively and achieve improved estimation accuracy with limited pilot resources. As the effective coherence interval becomes shorter, the covariance information becomes less reliable and the corresponding semi-blind gain decreases. In this regime, increasing the pilot ratio partially compensates for the degradation by providing a more accurate pilot-based channel estimate. These results characterize the practical operating regime of the proposed estimator under different channel coherence conditions and pilot-resource constraints.

\begin{figure}[t]
\centering
\includegraphics[width=0.98\linewidth]{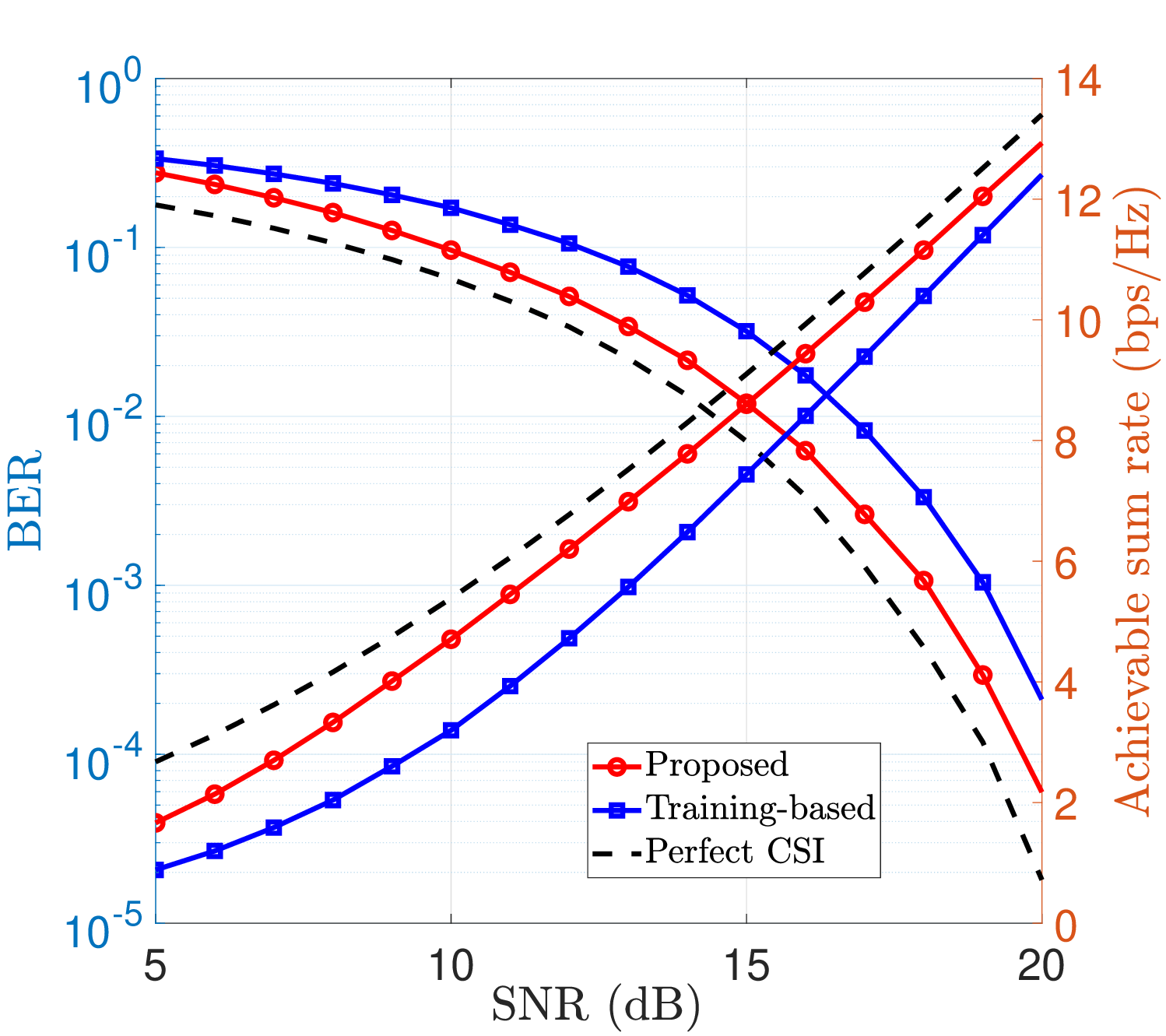}
\caption{BER and achievable sum rate versus SNR with $K=3$, $\alpha=1/4$, $\beta=1/16$, and $N=256$.}
\label{fig_SNR_BER_rate}
\end{figure}

\begin{figure}[t]
\centering
\includegraphics[width=1.0\linewidth]{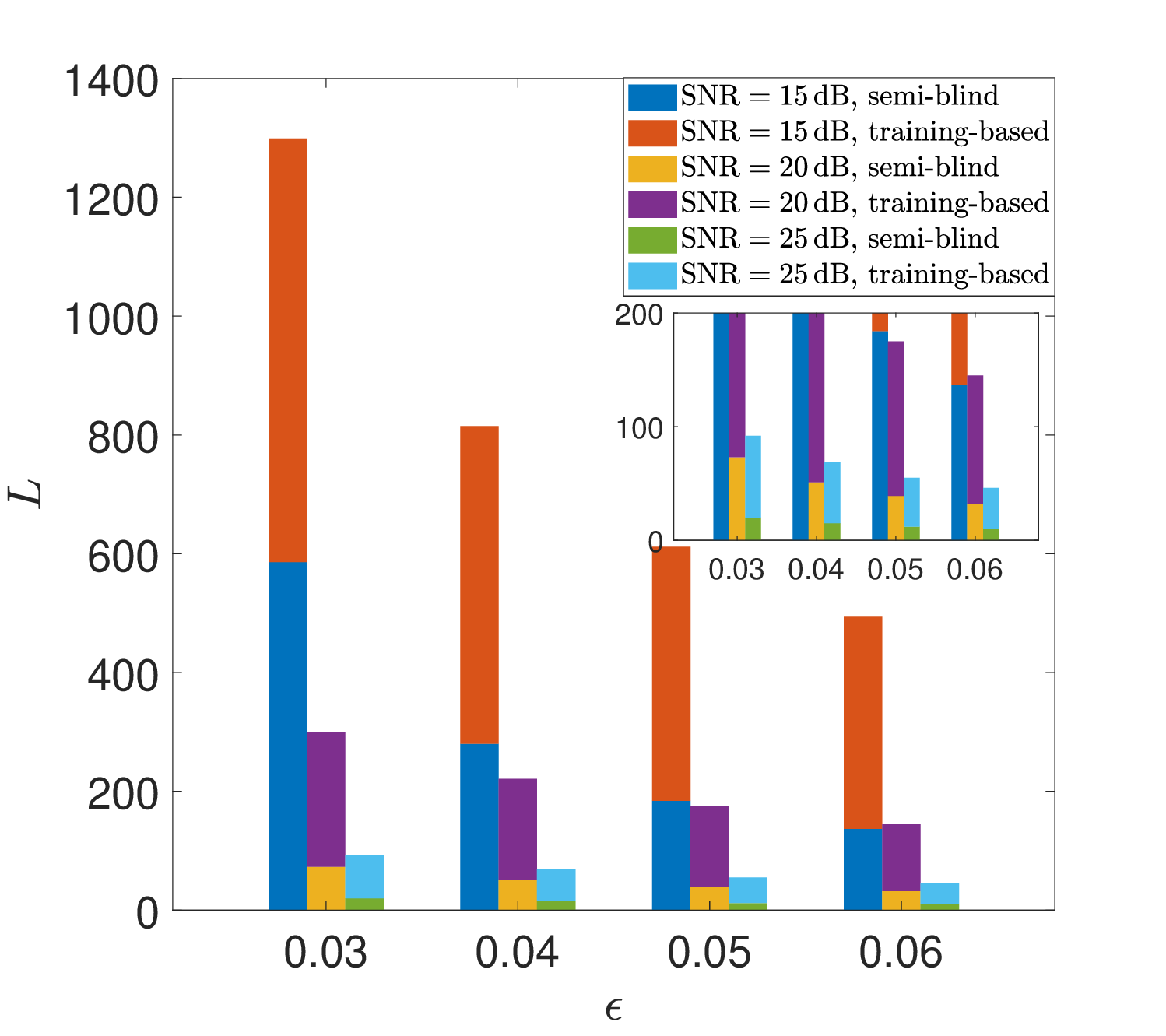}
\caption{The length $L$ of the training sequences versus $\epsilon$ with $K=3$, $\alpha=1/2$, $N=512$, and $\mathrm{SNR}=15$\,dB.}
\label{fig_L_required}
\end{figure}

\subsection{System-Level Performance Evaluation}
The preceding results evaluate the proposed estimator primarily in terms of channel estimation accuracy. To further assess whether the improved estimation accuracy translates into system-level communication gains, we evaluate the bit error rate (BER) and achievable sum rate using the estimated CSI. The channel estimates obtained by the proposed semi-blind estimator and the training-based estimator are employed for uplink minimum mean-square error (MMSE) data detection, while perfect CSI is included as a reference. The system parameters are set to $N=256$, $K=3$, $\alpha=1/4$, and $\beta=1/16$.

Fig.~\ref{fig_SNR_BER_rate} shows the BER and achievable sum rate versus SNR. In terms of BER, the proposed estimator consistently outperforms the training-based estimator over the entire considered SNR range and progressively approaches the perfect-CSI performance as the SNR increases. The performance gain becomes more pronounced in the moderate-to-high SNR regime, where residual channel estimation errors have a greater impact on data detection. This result confirms that the MSE improvement achieved by the proposed estimator translates into enhanced detection reliability.

For the achievable-rate evaluation, the sum rate is computed from the user SINRs achieved by the MMSE detector and scaled by the data-transmission fraction $1-\beta$ to account for the pilot overhead. As shown in Fig.~\ref{fig_SNR_BER_rate}, the proposed estimator consistently achieves a higher achievable sum rate than the training-based estimator over the entire considered SNR range and progressively approaches the perfect-CSI reference as the SNR increases. The improved channel estimation accuracy reduces the performance degradation caused by imperfect CSI in MMSE detection, thereby yielding higher spectral efficiency. These results demonstrate that the estimation gain provided by the proposed semi-blind estimator translates into improvements in both detection reliability and spectral efficiency.

\subsection{Evaluation of Pilot Requirements for a Target MSE}
Finally, we demonstrate the practical application of the derived MSE expressions. Given a target MSE value $\epsilon$, these expressions can be used to determine the number of pilot symbols required to achieve the desired accuracy for both training-based and semi-blind techniques, with all other system parameters $N$, $M$, and $K$ held constant.

For the training-based scheme, only the pilot symbols are utilized for channel estimation, while the symbol vectors received during data transmission are discarded. Based on (\ref{channel vector}) and (\ref{G_training}), the estimation error is given by
\begin{align}
    \hat{\mathbf{G}}_{\mathrm{training}} - \mathbf{G} = \frac{1}{aL}\mathbf{V}_p\mathbf{S}_p^H, 
\end{align}
Then,  the corresponding MSE is given as 
\begin{align}
    \mathrm{MSE}_{\mathrm{training}} = \frac{1}{K}\mathbb{E}\{\|\hat{\mathbf{G}}_{\mathrm{training}}-\mathbf{G}\|_F^2\} = \frac{\sigma_v^2M}{aL}.
\end{align}
This expression along with \eqref{eq:MSE_lambda} enable us to determine the number of pilot symbols $L$ required to meet a target MSE level $\epsilon$ for the training-based method and the proposed method. 
Figure~\ref{fig_L_required} depicts the required pilot length $L$ for different target MSE values, with the system parameters set to $N = 512$, $\alpha = 1/2$, $K = 3$, and $\mathrm{SNR}=15$\,dB.  As expected, with the increase of higher target MSE values, the number of required pilots decreases for both approaches. Notably, the semi-blind method requires fewer than half the pilot symbols compared to the training-based approach to achieve the same target MSE, highlighting its superior efficiency in pilot utilization.

\par

\section{Conclusion}
\label{sec_con}
In this paper, we proposed a semi-blind channel estimation method tailored for high-dimensional settings, where the number of unknown data symbols, training symbols, data symbols, and BS antennas grow large simultaneously while the number of users remains fixed. The estimator is formulated as the solution of a quadratic optimization problem whose objective function depends on a regularization parameter that properly weights the blind subspace criterion and the training-based criterion, thereby enabling the estimator to jointly exploit the subspace structure of the received data and the information conveyed by the pilot symbols.
Using RMT, we derived an asymptotic characterization of the MSE of the proposed estimator and obtained a consistent estimator of the optimal regularization parameter. This optimal parameter achieved the right balance between the blind and training-based components, yielding a simple, closed-form, and computationally efficient expression for the semi-blind channel estimator. Our method treated the channel as a deterministic but unknown quantity and did not rely on any underlying statistical distribution, which made it broadly applicable in realistic propagation conditions. To illustrate its practical relevance in realistic NTN environments, we applied the proposed estimator to NTN channel models. We showed that the resulting performance consistently outperformed existing approaches, such as the EM-based semi-blind algorithm, while achieving performance close to the CRB. These results highlight the potential of the proposed framework for reliable channel acquisition in future high-dimensional wireless systems operating over dynamic channels and advanced waveform receivers.

\appendices

\section{Proof of Theorem~\ref{theorem_MSE}}\label{proof_MSE}
To establish the uniform convergence of \(|{\rm MSE}(\lambda)-\overline{\rm MSE}(\lambda)|\) to zero, we start by expanding the difference as in~\eqref{diff_expansion}. Applying the triangle inequality to this decomposition yields the upper bounds stated in \eqref{uniform_MSE}. We then analyze the asymptotic behavior of each of the 
three terms in~\eqref{uniform_MSE} and show that they converge to zero 
uniformly for all \(\lambda \in [0,1]\).
\begin{figure*}
\begin{align}
    &|\mathrm{MSE}(\lambda) - \overline{\mathrm{MSE}}(\lambda)|\notag\\
    &= \left|\frac{(1-\lambda)^2}{K}\sum_{k=1}^K\Big(\mathbf{g}_k^H\mathbf{g}_k-\frac{\sigma_v^2}{P_s}t_k\Big) +\frac{(1-\lambda)^2}{K}\sum_{k=1}^K\Big(\sigma_v^2z_kt_k-\sum_{i=1}^K\mathbf{g}_k^H\hat{\mathbf{u}}_{si}\hat{\mathbf{u}}_{si}^H\mathbf{g}_k\Big)+ \frac{\sigma_v^2(\lambda^2M+(1-\lambda^2)K)}{aL} - \frac{\sigma_v^2\lambda^2\alpha}{a\beta}\right| \label{diff_expansion}\\
    &\leq \frac{(1-\lambda)^2}{K}\left|\sum_{k=1}^K\Big(\mathbf{g}_k^H\mathbf{g}_k-\frac{\sigma_v^2}{P_s}t_k\Big)\right| + \frac{(1-\lambda)^2}{K}\left|\sum_{k=1}^K\Big(\sum_{i=1}^K\mathbf{g}_k^H\hat{\mathbf{u}}_{si}\hat{\mathbf{u}}_{si}^H\mathbf{g}_k-\frac{\sigma_v^2}{P_s}z_kt_k\Big)\right| + \left|\frac{\sigma_v^2(\lambda^2M+(1-\lambda^2)K)}{aL} - \frac{\sigma_v^2\lambda^2\alpha}{a\beta}\right|. \label{uniform_MSE}
\end{align}
\hrulefill
\vspace*{10pt}
\end{figure*}

1) 
The first term is exactly zero since:
\begin{align}
    \sum_{k=1}^K\mathbf{g}_k^H\mathbf{g}_k = \mathrm{tr}(\mathbf{G}^H\mathbf{G}) = \mathrm{tr}(\mathbf{GG}^H) = \frac{\sigma_v^2}{P_s}\sum_{k=1}^Kt_k.
\end{align}

2)  Using the result in ~\cite{couillet2011random,yang2018high}, we have
\begin{align}\label{As_z_i}
    \left|\mathbf{g}_k^H\hat{\mathbf{u}}_{si}\hat{\mathbf{u}}_{si}^H\mathbf{g}_k -z_i\mathbf{g}_k^H\mathbf{u}_{si}\mathbf{u}_{si}^H\mathbf{g}_k\right| \overset{a.s.}{\longrightarrow} 0, \,\, i=1,\ldots,K,
\end{align}
from which we control the second term in \eqref{uniform_MSE} as:
\begin{align}
    &\left|\sum_{k=1}^K\left(\sum_{i=1}^K\mathbf{g}_k^H\hat{\mathbf{u}}_{si}\hat{\mathbf{u}}_{si}^H\mathbf{g}_k-z_kt_k\right)\right| \notag\\
    &\qquad=  \left|\sum_{k=1}^K\sum_{i=1}^K(\mathbf{g}_k^H\hat{\mathbf{u}}_{si}\hat{\mathbf{u}}_{si}^H\mathbf{g}_k-\frac{\sigma_v^2}{P_s}z_i\mathbf{g}_k^H\mathbf{u}_{si}\mathbf{u}_{si}^H\mathbf{g}_k)\right| \notag\\
    &\qquad\leq \sum_{k=1}^K\sum_{i=1}^K|\mathbf{g}_k^H\hat{\mathbf{u}}_{si}\hat{\mathbf{u}}_{si}^H\mathbf{g}_k-\frac{\sigma_v^2}{P_s}z_i\mathbf{g}_k^H\mathbf{u}_{si}\mathbf{u}_{si}^H\mathbf{g}_k|
    \overset{a.s.}{\longrightarrow} 0,
\end{align}
where in the second equality we used the fact that:
\begin{align}
    \sum_{k=1}^K\sum_{i=1}^Kz_i\mathbf{g}_k^H\mathbf{u}_{si}\mathbf{u}_{si}^H\mathbf{g}_k =& \sum_{i=1}^Kz_i\mathbf{u}_{si}^H\left(\sum_{k=1}^K\mathbf{g}_k\mathbf{g}_k^H\right)\mathbf{u}_{si} \notag\\
    =& \sum_{i=1}^Kz_i\mathbf{u}_{si}^H\mathbf{G}\mathbf{G}^H\mathbf{u}_{si} \notag\\
    =& \frac{\sigma_v^2}{P_s}\sum_{k=1}^Kz_kt_k.
\end{align}

3) As $M/N\rightarrow\alpha$, $L/N\rightarrow\beta$ with fixed $K$, $M/L\rightarrow\alpha/\beta$, and $K/L\rightarrow0$, which yields
\begin{align}
    \left|\frac{\sigma_v^2(\lambda^2M+(1-\lambda^2)K)}{aL} - \frac{\sigma_v^2\lambda^2\alpha}{a\beta}\right| \xrightarrow[N\to\infty]{a.s.} 0. 
\end{align}
Combining the asymptotic results for all three terms, and the fact that $\mathrm{MSE}(\lambda)$ is an analytic function with respect to $\lambda$, we conclude that
\begin{align}
    |\mathrm{MSE}(\lambda) - \overline{\mathrm{MSE}}(\lambda)|\xrightarrow[N\to\infty]{a.s.} 0.
\end{align}
which establishes the uniform convergence of $\mathrm{MSE}(\lambda)$ over $\lambda\in[0,1]$. 

\section{Proof of Theorem~\ref{theorem_lambda}}\label{proof_lambda}
Using the results in ~\cite{couillet2011random,yao2015sample,yang2018high}, for $t_i>\sqrt{c}$, $\eta_i$ satisfies the following convergence:
\begin{align}\label{lambda_t_i}
    \frac{\eta_i}{\sigma_v^2}\overset{a.s.}{\longrightarrow} 1+t_i+\frac{c(1+t_i)}{t_i},
\end{align}
By solving the equation $\frac{\eta_i}{\sigma_v^2}= 1+\hat{t}_i+\frac{c(1+\hat{t}_i)}{\hat{t}_i}$, we obtain consistent estimators of $t_i$, i.e., 
\begin{align}
    |\hat{t}_i-t_i|\overset{a.s.}{\longrightarrow} 0, i = 1,\ldots,K,
\end{align}
where 
\begin{align}
    \hat{t}_i = \frac{\eta_i/\sigma_v^2+1-c+\sqrt{(\eta_i/\sigma_v^2+1-c)^2-4\eta_i/\sigma_v^2}}{2} -1. 
\end{align}
Finally, using $\hat{t}_i$, a consistent estimator of $z_i$ is  $$\hat{z}_i=\frac{1-c/\hat{t}_i^2}{1+c/\hat{t}_i}.
$$
Therefore, we obtain consistent estimators $\hat{\lambda}^{\ast}$ of the asymptotically-optimal $\bar{\lambda}^{\ast}$ by substituting $\hat{z}_i$ and $\hat{t}_i$ for $z_i$ and $t_i$, respectively.

\section{Proof of Theorem~\ref{theorem_MSE_lambda}}\label{proof_MSE_lambda}
To begin with, we apply triangle inequality to obtain
\begin{align}
    |\mathrm{MSE}(\hat{\lambda}^{\ast}) - \overline{\mathrm{MSE}}(\bar{\lambda}^{\ast})| 
    \leq& |\mathrm{MSE}(\hat{\lambda}^{\ast}) - \overline{\mathrm{MSE}}(\hat{\lambda}^{\ast})| \notag\\
    &+ |\overline{\mathrm{MSE}}(\hat{\lambda}^{\ast}) - \overline{\mathrm{MSE}}(\bar{\lambda}^{\ast})|. 
\end{align}
Then we analyze the two terms on the right-hand side separately. From Theorem~\ref{theorem_MSE}, we have
\begin{align}
    &|\mathrm{MSE}(\hat{\lambda}^{\ast}) - \overline{\mathrm{MSE}}(\hat{\lambda}^{\ast})| \notag\\
    &\qquad\qquad\quad\leq \max_{\lambda\in[0,1]}|\mathrm{MSE}(\lambda) - \overline{\mathrm{MSE}}(\lambda)|\xrightarrow[N\to\infty]{a.s.} 0. 
\end{align}
Then from the expression of derivative in (\ref{derivative of MSE}), we can obtain its upper-bound as 
\begin{align}
    \frac{\partial\overline{\mathrm{MSE}}(\lambda)}{\partial\lambda}\leq \frac{2\sigma_v^2\alpha}{a\beta} \triangleq Q. 
\end{align}
This implies that $\overline{\mathrm{MSE}}(\lambda)$ is Lipschitz continuous with constant $Q$ on $[0,1]$ \cite{lebl2009basic,li2012consensus,zuhlke2025adversarial}. that is 
\begin{align}
    |\overline{\mathrm{MSE}}(\hat{\lambda}^{\ast}) - \overline{\mathrm{MSE}}(\bar{\lambda}^{\ast})| \leq Q|\hat{\lambda}^{\ast}-\bar{\lambda}^{\ast}|\overset{a.s.}{\longrightarrow} 0. 
\end{align}
 Combining the two convergence results above, we conclude that $ |\mathrm{MSE}(\hat{\lambda}^{\ast}) - \overline{\mathrm{MSE}}(\bar{\lambda}^{\ast})|\overset{a.s.}{\longrightarrow} 0$.

\bibliographystyle{IEEEtran}
\bibliography{references}

\end{document}